\documentclass[aps,preprint]{revtex4-1}
\usepackage{graphicx}
\usepackage{amsmath}
\usepackage{makeidx}
\usepackage{amsfonts}
\usepackage{amssymb}
\usepackage{dsfont}
\usepackage{color,soul}

\begin{document}

\title{State assignment problem in systems biology and medicine: on the importance of state interaction network topology}

\author{Abolfazl Ramezanpour$^{a,b}$, Alireza Mashaghi$^{a,c,\ast}$}
\affiliation{$^a$Leiden Academic Centre for Drug Research, Faculty of Mathematics and Natural Sciences, Leiden University, Leiden, The Netherlands}
\affiliation{$^b$Department of Physics, University of Neyshabur, Neyshabur, Iran}
\affiliation{$^c$Harvard Medical School, Harvard University, Boston, Massachusetts, USA}
\affiliation{$^\ast$Correspondence to a.mashaghi.tabari@lacdr.leidenuniv.nl}
\date{\today}

\begin{abstract}
A fundamental problem in medicine and biology is to assign states, e.g. healthy or diseased, to cells, organs or individuals. State assignment or making a diagnosis is often a nontrivial and challenging process and, with the  advent of omics technologies, the diagnostic challenge is becoming more and more serious. The challenge lies not only in the increasing number of measured properties and dynamics of the system (e.g. cell or human body) but also in the co-evolution of multiple states and overlapping properties, and degeneracy of states. We develop, from first principles, a generic rational framework for state assignment in cell biology and medicine, and demonstrate its applicability with a few simple theoretical case studies from medical diagnostics. We show how disease-related statistical information can be used to build a comprehensive model that includes the relevant dependencies between clinical and laboratory findings (signs) and diseases. In particular, we include disease-disease and sign-sign interactions. We then study how one can infer the probability of a disease in a patient with given signs. We perform comparative analysis with simple benchmark models to check the performances of our models. This first principles approach, as we show, enables the construction of consensus diagnostic flow charts and facilitates the early diagnosis of disease. Additionally, we envision that our approach will find applications in systems biology, and in particular, in characterizing the phenome via the metabolome, the proteome, the transcriptome, and the genome. 
\end{abstract}

\maketitle

\section{Introduction}\label{S0}

Human body as a whole or in part may adopt various states, like a Rubik's Cube. Homeostatic mechanisms, medical interventions and aging all involve evolution from certain body states to others. Similarly, evolution of states is commonly seen in cells that constitute our bodies. Immune cells for example can manifest substantial plasticity and develop into distinct phenotypes with different functions \cite{DuPage2016, Holzel2016}. Identifying dysfunctional states in cells is in many ways similar to identifying diseases in organisms and is confronted by similar difficulties. State assignment, as we describe below, is often a nontrivial and challenging process and in many cases it is hard to “diagnose” the state of a cell or conditions of a patient. The progress in systems biology and availability of large data has made diagnostics even more challenging. For instance, mass cytometric analysis of immune cells has led to identification of many new cell subtypes (states) \cite{Saeys2016}. Metabolomic analysis of cells and body fluids revealed a large number of biomarkers and disease subtypes \cite{Patti2012}. There is a huge need in the fields of cell biology, immunology, clinical sciences and pharmaceutical sciences for approaches to identify states, assigning states and characterizing co-emerging or co-existing states. Moreover, it is often important to be able to identify emerging states even before they are fully evolved. From physics point of view, it is interesting to yield a generic understanding of the state assignment problem in cell biology or medicine (although specific details might be crucial in each context in practice). Without loss of generality, in what follows we focus on medical diagnostics and draw a simple picture that captures many generic aspects of assignment problems in medicine and systems cell biology. 

Decision-making is at the heart of medicine. Decisions are made at various stages in clinical practice, particularly during diagnostic investigations and when assigning the findings to a disease \cite{Harrison-2015, Current-2016}. Diagnostic strategies are typically available in the form of clinical algorithms and flow charts that define the sequence of actions to be taken to reach a diagnosis. The diagnosis itself is typically made based on consensus diagnostic criteria \cite{Lerner-2006, Coggon-2005}. In addition, there are a number of clinical decision support systems and software systems that are used to assign findings (symptoms and signs) to disease conditions. The most commonly used technologies are WebMD Symptom Checker, Isabel Symptom Checker, DXplain and Internist\cite{Support-2014}. These algorithms compute the most likely disease that is associated with a given set of findings by using only a small part of the existing probabilistic data on findings and diseases. Internist, which is one of the most sophisticated systems, relies on two parameters, the probability of a finding given a disease and the probability of a disease given a finding \cite{internist}. These technologies inform us if a patient satisfies the criteria of a disease but do not provide guidance on how to approach a patient and mostly ignore the interactions between diseases.

Currently, we lack a solid conceptual framework for medical diagnostics. As a consequence, there is no consensus on the diagnostic flow charts available today, and clinicians differ widely in their approaches to patients. Here, we take a step towards solving this problem by formulating first principles medical diagnostics. We evaluate the performance of the platform and discuss how one can optimize it. Using simple theoretical examples, we show how including relevant statistical data and often-ignored inherent disease-disease linkages significantly reduces diagnostic errors and mismanagement and enables the early diagnosis of disease.   


The problem of associating a subset of observed signs (clinical signs/symptoms and laboratory data) with specific diseases was easy if we could assume the findings originate from a single disease, we had clear demonstrations for the diseases, and inter-sign and inter-disease interactions were negligible. In practice, however, we typically have no certain relationships that connect signs to diseases, and one often must address interference effects of multiple diseases; in the early stages of a disease, we do not even have sufficient findings to make a definite decision  \cite{Heckerman-1987,HBH-ar-1988,shortlife}. There are a number of studies that have attempted to quantify such dependencies under uncertainty and obtain estimations for the likelihood of diseases given a subset of findings \cite{Adams-1976,internist,mycin,qmr,dxplain,Spiegelhalter,Bankowitz}. An essential simplifying assumption in these studies was that only one disease is behind the findings (exclusive diseases assumption). Among recent developments, we should mention Bayesian belief networks, which provide a probabilistic framework to study sign-disease dependencies \cite{Heckerman-1990,internist-1,Heckerman-1,Nikovski-ieee-2000}. These models are represented by tables of conditional probabilities that show how the state of a node (sign or disease) variable in an acyclic directed graph depends on the state of the parent variables. Here, it is usually assumed that the signs are conditionally independent of one another given a disease hypothesis and that diseases are independent of one another after marginalizing over the sign variables (marginally independent diseases)\cite{internist-1}. In other words, there exist no causal dependencies or interactions (directed links in the acyclic graph) that connect the signs/diseases. In this study, however, we shall pursue a more principled approach to highlight the significance of direct disease-disease and sign-sign interactions (dependencies). Evidences are rapidly growing to support the existence of such interactions \cite{Przulj2016, Hamaneh2015, zitnik2013, LiuWu2015, Suratanee2015, Gustafsson2014, LiuTseng2014, SunGoncalves2014}. Our approach is of course computationally more expensive than the previous approaches, but it shows how different types of interactions could be helpful in the course of diagnosis. Additionally, because of recent developments in related fields \cite{Jordan,MM-book-2009,ABRZ-prl-2011,T-jstat-2012}, we now have the necessary concepts and tools to address difficulties in more sophisticated (realistic) problems of this type. This study does not involve usage of real medical data, which is by the way fundamentally incomplete at this moment for such modeling; however, it provides a rationale as to why certain often-neglected statistical information and medical data can be useful in diagnosis and demonstrates that investments in collecting such data will likely pay off.  

\section{Problem statement}\label{S01}
Consider a set of $N_D$ binary variables $\mathbf{D}=\{D_{a}=0,1: a=1,\dots,N_D\}$, where $D_{a}=0,1$ shows the absence or presence of disease $a$. We have another set of $N_S$ binary variables $\mathbf{S}=\{S_i=\pm 1: i=1,\dots,N_S\}$ to show the values of sign (symptom) variables. 

Suppose we have the conditional probability of symptoms given a disease hypothesis, $P(\mathbf{S}|\mathbf{D})$, and prior probability of diseases $P_0(\mathbf{D})$. Then, the joint probability distribution of sign and disease variables reads as $P(\mathbf{S};\mathbf{D})\equiv P(\mathbf{S}|\mathbf{D})P_0(\mathbf{D})$. We shall assume, for simplicity, that the probability distributions describe the stationary state of the variables. The distributions may be subject to environmental and evolutionary changes and may also change in the course of the disease. Here, we limit our study to the time scales that are smaller than the dynamical time scale of the model and leave the temporal dynamics for a future study. In addition, we assume that we are given sufficient statistical data, e.g., the true marginal probabilities $P_{true}(S_i,S_j|\mathbf{D})$, to reconstruct simple models of the true probability distribution \cite{Jaynes-book-2003}. This is indeed the first part of our study: In Sec. \ref{S11}, we propose statistical models of sign and disease variables, and employ efficient learning algorithms to compute the model parameters, given the appropriate data. Fortunately, recent advances in machine learning and inference techniques enable us to work with models that involve very large number of variables \cite{KR-nc-1998,T-pre-1998,SBSB-nature-2006,CLM-pnas-2009,WWSHH-pnas-2009,RAH-cn-2009,BBPWZ-bb-2010,NB-prl-2012,T-jstat-2012}.

Let us assume that a subset $O=\{i_1,i_2,\dots,i_{N_O}\}$ of the sign variables has been observed with values    
$\mathbf{S}^o$, and size $N_O=|O|$. We will use $U$ for the remaining subset of unobserved signs with values which are denoted by $\mathbf{S}^u$. 
Then, the likelihood of disease variables given the observed signs is:  
\begin{align}
\mathcal{L}(\mathbf{D}|\mathbf{S}^o)\equiv \sum_{\mathbf{S}^u} P(\mathbf{S};\mathbf{D}).
\end{align}  
The most likely diseases are obtained by maximizing the above likelihood:
\begin{align}
\mathbf{D}_{ML}=\arg \max_{\mathbf{D}} \log \mathcal{L}(\mathbf{D}|\mathbf{S}^o).
\end{align}  
Here, we are interested in the posterior probability marginals $P(D_a=0,1)$ of the disease variables.  
The marginal probability $P(S_j=\pm 1)$ of an unobserved sign, and the most likely signs, are obtained from the following distribution:
\begin{align}
\mathcal{M}(\mathbf{S}^u|\mathbf{S}^o)\equiv \sum_{\mathbf{D}} P(\mathbf{S};\mathbf{D}),\hskip1cm
\mathbf{S}_{ML}=\arg \max_{\mathbf{S}^u} \log \mathcal{M}(\mathbf{S}^u|\mathbf{S}^o).
\end{align}
The main task in the second part of our study is computation of the sign and disease marginal probabilities
\begin{align}
P(D_{a}) \propto \sum_{\{D_b:b \neq a\}} \mathcal{L}(\mathbf{D}|\mathbf{S}^o),\hskip1cm 
P(S_j) \propto \sum_{\{S_k:k \in U\setminus j\}} \mathcal{M}(\mathbf{S}^u|\mathbf{S}^o) \hskip1cm j \in U.
\end{align}  
In general, computing the exact values of these marginals is a hard problem. However, one can find highly accurate approximation methods developed in the artificial intelligence and statistical physics communities to address such computationally difficult problems \cite{Pearl,KFL-inform-2001,MPZ-science-2002,YFW-ai-2003,Jordan,MM-book-2009}. In Sec. \ref{S12}, we propose an approximate message-passing algorithm for inferring the above information in a large-scale problem.

Finally, the last and main part of our study is devoted to the problem of choosing a finite sequence of unobserved signs for observation, which maximizes an appropriate objective functional of the sequence of observations. In principle, the objective function should be designed to approach the right diagnosis in a small number of observations. To this end, we assign larger values to the objective function if the observations result to larger polarization in the disease probabilities; obviously, it is easier to decide if disease $a$ is present or not when the marginal probability $P(D_a)$ is closer to $0$ or $1$ (more polarized). Computing such an objective functional of the disease probabilities for a given sequence of observations is not an easy task. We have to consider also the stochastic nature of the observations; we know the sign probabilities $P(S_j)$, but, we do not know a priori the value $S_j$ of an unobserved sign, which is chosen for observation. To take into account this uncertainty, we shall work with an objective function which is averaged over the possible outcomes of the observation. More precisely, the above diagnosis problem is a multistage stochastic optimization problem, a subject that has been extensively studied in the optimization community \cite{BL-book-1997,KSH-siam-2002,HS-book-2003,ABRZ-prl-2011}.

Suppose we are to observe $T\le N_S-N_O$ signs with an specified order $O_T\equiv \{j_1,\dots,j_T\}$; there are $(N_S-N_O)!/(T!(N_S-N_O-T)!)$ different ways of choosing $T$ signs from $N_S-N_O$ ones, and $T!$ different orderings of the selected signs to identify such a sequence of observations. 
Therefore, the number of possible sequences grows exponentially with $T$. Add to this the computational complexity of working with an objective functional of the sequence of observations, which has to be also averaged over the stochastic outcomes of the observations. In Sec. \ref{S13}, we present simple heuristic and greedy algorithms to address the above problem, and leave a detailed study of the multistage stochastic optimization problem for future.

\section{Results}\label{S1}
A complete description of a collection of stochastic variables, like the sign and disease variables, is provided by the joint probability distribution of the variables $P(\mathbf{S};\mathbf{D})$. Having the probability distribution (model) that describes a system of interacting variables does not, however, mean that one can readily extract useful statistical information from the model. In fact, both the model construction and the task of extracting information from the model are computationally hard, with computation times that in the worst cases grow exponentially with the number of involved variables \cite{GJ-book-1979,cooper-ai-1990,Henrion-1990,HGC-ml-1995,C-learning-1996}. In the following, we address the above sub-problems in addition to the main problem of optimizing an appropriate objective functional of observations, which are made during the course of diagnosis.          

\begin{figure}
\includegraphics[width=8cm]{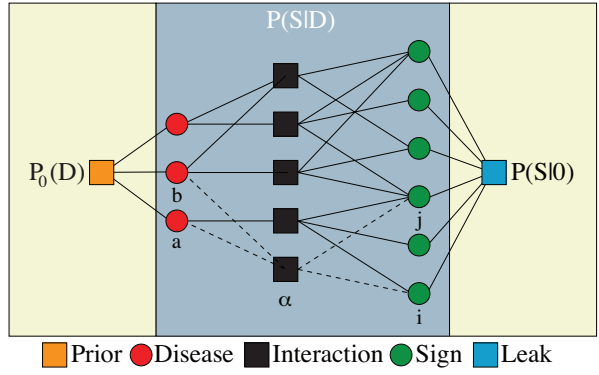} 
\caption{The interaction graph of disease variables (left circles) and sign variables (right circles) related by $M_a$ one-disease and $M_{ab}$ two-disease interaction factors (middle squares) in addition to interactions induced by the leak probability (right square) and the prior probability of diseases (left square). In general, an interaction factor $\alpha=a,ab$ is connected to $k_{\alpha}$ signs and $l_{\alpha}$ diseases.}\label{dsf}
\end{figure}

\subsection{Learning the model: Maximum entropy principle}\label{S11}
Let us assume that we are given the marginal probabilities $P_{true}(S_i,S_j|\mathbf{D})$ of the true conditional probability $P_{true}(\mathbf{S}|\mathbf{D})$. Then, we use the maximum entropy principle to construct an appropriate model $P(\mathbf{S};\mathbf{D})\equiv P(\mathbf{S}|\mathbf{D})P_0(\mathbf{D})$ of the sign and disease variables \cite{LBCMF-pnas-2006,BR-arxiv-2007,BMV-jpc-2010}. Here, $P_0(\mathbf{D})$ is the prior probability of the diseases depending on the age, gender, and other characteristics. In the following, we simply take a product prior distribution, $P_0(\mathbf{D})=\prod_{a} P_{0}(D_{a})$.
In the absence of any prior information, the above probability distribution is uniform. The conditional probability $P(\mathbf{S}|\mathbf{D})$ represents all the sign/disease interactions that are allowed by the maximum entropy principle,
\begin{align}
P(\mathbf{S}|\mathbf{D}) = \frac{1}{Z(\mathbf{D})}\exp\left(\sum_i h_i(\mathbf{D})S_i+\sum_{i<j}J_{ij}(\mathbf{D})S_iS_j \right).
\end{align}  
Here $Z(\mathbf{D})$ is the normalization (or partition) function.

In practice, we are given only a small subset of the conditional probabilities, for instance, $P_{true}(S_i,S_j|\mathrm{only} D_a)$ and $P_{true}(S_i,S_j|\mathrm{only} D_a,D_b)$. The former is the probability that signs $i$ and $j$ take values $(S_i=\pm 1,S_j=\pm 1)$ conditioned to the presence of disease $a$ and the absence of all other diseases. The latter conditional probabilities are defined similarly. Therefore, we have to consider only interactions between a small number of disease variables. To this end, we expand the model parameters, 
\begin{align}
h_i(\mathbf{D}) &=h_i^0+\sum_a h_i^a D_a+\sum_{a<b}h_i^{ab}D_aD_b+\cdots,\\
J_{ij}(\mathbf{D}) &=J_{ij}^0+\sum_a J_{ij}^a D_a+\sum_{a<b}J_{ij}^{ab}D_aD_b+\cdots, 
\end{align}  
and keep only the leading terms of the expansion.
Putting all together, given the above information, we rewrite 
\begin{align}\label{CPSD}
P(\mathbf{S}|\mathbf{D})=\frac{1}{Z(\mathbf{D})}\phi_0(\mathbf{S}) \times \prod_{a} \phi_{a}(\mathbf{S}|D_a)\times  \prod_{a < b} \phi_{ab}(\mathbf{S}|D_a,D_b).
\end{align}
Here, $\phi_0$ is responsible for the leak probabilities $P(\mathbf{S}|\mathrm{nodisease})$, to account for the missing disease information and other sources of error \cite{internist-1,Nikovski-ieee-2000}. In the following, we assume that local sign fields are sufficient to produce an accurate representation of the leak probabilities, i.e., $\phi_0=\exp(\sum_i K_i^0S_i)$ . The other interaction factors, $\phi_{a}$ and $\phi_{ab}$, are present only if the associated diseases are present; they are written in terms of local sign fields and two-sign interactions:
\begin{align}\label{phis}
\phi_a &= \exp(D_a[\sum_i K_i^{a}S_i+\sum_{i<j} K_{ij}^{a}S_iS_j]),\\
\phi_{ab} &= \exp(D_{a}D_{b}[\sum_i K_i^{ab} S_i+\sum_{i<j} K_{ij}^{ab} S_iS_j]).
\end{align}

Figure \ref{dsf} shows a graphical representation of the model, which has $M_a$ one-disease and $M_{ab}$ two-disease interaction factors, each of which is connected to $k_a$ and $k_{ab}$ sign variables, respectively. From the above model, we obtain the simpler one-disease-one-sign ($D1S1$) model in which we have only the one-disease factors (i.e., $M_{ab}=0$) and local sign fields (i.e., $K_{ij}^a=0$). In a two-disease-one-sign model ($D2S1$), we have both the one- and two-disease factors, but only the local sign fields. In the same way, we define the one-disease-two-sign ($D1S2$) and two-disease-two-sign ($D2S2$) models.
In the following, unless otherwise mentioned, we shall work with the fully connected graphs with parameters: $M_a=N_D, k_a=N_S$ for the $D1S1$ and $D1S2$ models, and $M_a=N_D, M_{ab}=N_D(N_D-1)/2, k_a=k_{ab}=N_S$ for the $D2S1$ and $D2S2$ models. Moreover, the interaction factors in the $D1S2$ and $D2S2$ models include all the possible two-sign interactions in addition to the local sign fields.

To obtain the model parameters ($K_i^{0},K_i^{a,ab}$, and $K_{ij}^{a,ab}$), we start from the conditional marginals $P_{true}(S_i|\mathrm{nodisease})$. This information is sufficient to determine the couplings $K_i^0$ from the following consistency equations:
\begin{align}
P_{true}(S_i|\mathrm{nodisease}) = \sum_{\{S_j:j\neq i\}}P(\mathbf{S}|\mathbf{D}=\mathbf{0}) \hskip1cm \forall i.
\end{align}  
If we have $P_{true}(S_i,S_j|\mathrm{only}D_{a})$, then in principle we can find $K_{i}^{a}$ and $K_{ij}^{a}$ from similar consistency equations, assuming that we already know the $K_i^0$. Note that $P(S_i,S_j|\mathrm{only}D_{a})$ is different from $P(S_i,S_j|D_{a})$, which is conditioned only on the value of disease $a$. In the same way, having the $P_{true}(S_i,S_j|\mathrm{only}D_{a},D_{b})$ allow us to find the couplings $K_{i}^{ab}$ and $K_{ij}^{ab}$, and so on. In general, the problem of finding the couplings from the above conditional probabilities is computationally expensive. However, there are many efficient approximate methods that enable us to find good estimations for the above couplings given the above conditional probabilities \cite{KR-nc-1998,T-pre-1998,SBSB-nature-2006,CLM-pnas-2009,RAH-cn-2009,NB-prl-2012,T-jstat-2012}. The reader can find more details about the models in Appendix \ref{app-1}, where we provide a very simple learning algorithm, which is based on the Bethe approximation, for estimating the model parameters, given the above marginal probabilities.

\subsection{Computing the marginal probabilities: An approximate inference algorithm}\label{S12}
Let us consider a simple benchmark model to check the performances of the above constructed models. As the benchmark, we take the true conditional probability
\begin{align}\label{exp-model}
P_{true}(\mathbf{S}|\mathbf{D})=\frac{1}{Z_{true}(\mathbf{D})}e^{-H(\mathbf{S},\mathbf{S}^*(\mathbf{D}))},
\end{align} 
where $\mathbf{S}^*(\mathbf{D})$ gives the symptoms of hypothesis $\mathbf{D}$; we choose these symptoms randomly and uniformly from the space of sign variables.
Moreover, $H(\mathbf{S},\mathbf{S}^*(\mathbf{D})) \equiv \sum_{i=1}^{N_S}(S_i-S_i^*(\mathbf{D}))^2/4$ is the Hamming distance (number of different signs) of the two sign configurations. Note that there is no sign-sign interaction in the above true model. Therefore, given the true conditional marginals, we can exactly compute the model parameters, as described in Appendix \ref{app-1}.      

\begin{figure}
\includegraphics[width=8cm]{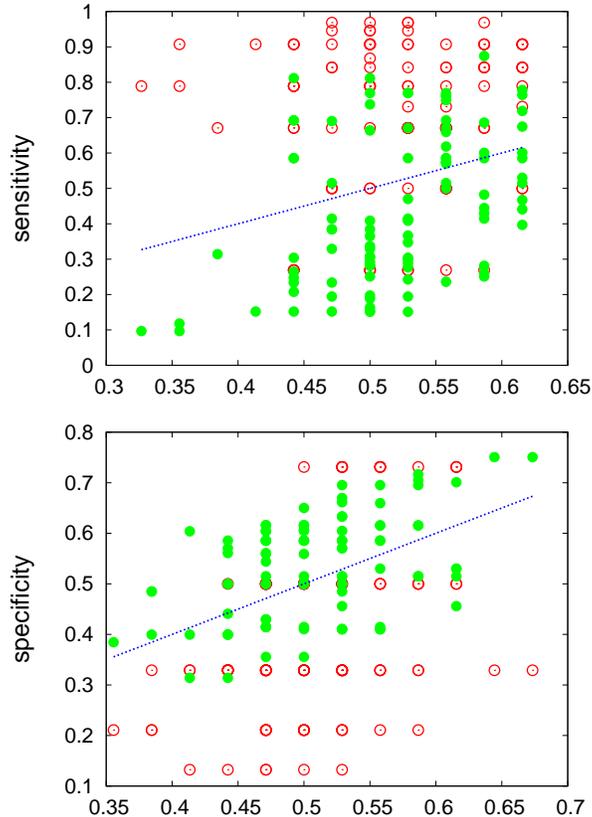} 
\caption{Sensitivity, $P(S_i=+1|D_a=1)$, and specificity, $P(S_i=-1|D_a=0)$, of the $D1S1$ (open circles) and $D2S1$ (filled circles) models versus the true values for a typical realization of the exponential true model. The model parameters are obtained from the conditional marginals of the true model. Here we have $N_D=5$ diseases, $N_S=20$ signs.}\label{Snp-1}
\end{figure}

For small numbers of sign/disease variables, we can use an exhaustive inference algorithm to compute the exact marginal probabilities. Figure \ref{Snp-1} displays the sensitivity $P(S_i=+1|D_a=1)$ and specificity $P(S_i=-1|D_a=0)$ of the diseases \cite{Nikovski-ieee-2000}, which have been obtained by the one-disease-one-sing ($D1S1$) and two-disease-one-sing ($D2S1$) models for a typical realization of the symptoms $\mathbf{S}^*(\mathbf{D})$ in the true model. As the figure shows, both the probabilities are in average closer to the true values in the $D2S1$ model. This computation is intended to exhibit the high impact of two-disease interactions on the behavior of the marginal probabilities. We will soon see that these large effects of interactions can indeed play a constructive role also in the process of diagnosis.

To infer the marginal probabilities of the models for larger number of sign/disease variables, we resort to the Bethe approximation and the Belief-Propagation algorithm \cite{KFL-inform-2001,MM-book-2009}. First, we suggest an approximate expression for the normalization function $Z(\mathbf{D})$, which appears in the denominator of the conditional probability distribution $P(\mathbf{S}|\mathbf{D})$. In words, we consider this non-negative function of the diseases to be a probability measure, and we approximate this measure by a factorized probability distribution, using its one- and two-variable marginal probabilities (see Appendix \ref{app-2}). This approximation enables us to employ an efficient message-passing algorithm such as belief propagation for computing statistical properties of the above models. As mentioned before, we shall assume that the prior probability $P_0(\mathbf{D})$ can also be written in an appropriate factorized form. The quality of our approximations depends very much on the structure of the interaction factors and the strengths of the associated couplings in the models. The Bethe approximation is exact for interaction graphs that have a tree structure. This approximation is also expected to work very well in sparsely connected graphs, in which the number of interaction factors ($M_a, M_{ab}$) and the number of signs associated with an interaction factor ($k_a, k_{ab}$) are small compared with the total number of sign variables. In Appendix \ref{app-2} we display the relative errors in the marginal signs/diseases probabilities that were obtained by the above approximate algorithm. The time complexity of our approximate inference algorithm grows linearly with the number of interaction factors and exponentially with the number of variables that are involved in such interactions; with $N_D=500, N_S=5000, M_a=500, M_{ab}=1000, k_a=10, k_{ab}=5$, the algorithm takes approximately one minute of CPU time on a standard PC to compute the local marginals. We recall that the INTERNIST algorithm works with $534$ diseases and approximately $4040$ signs (or manifestations), with $40740$ directed links that connect the diseases to the signs \cite{internist-1}.

\subsection{Optimization of the diagnosis process: A stochastic optimization problem}\label{S13}
Suppose that we know the results of $N_O$ observations (medical tests), and we choose another unobserved sign $j\in U$ for observation. To measure the performance of our decision, we may compute deviation of the disease probabilities from the neutral values (or "disease polarization") after the observation:          
\begin{align}
DP(j)\equiv  \left( \frac{1}{N_D} \sum_a \left(P(D_{a}=1)-\frac{1}{2}\right)^2 \right)^{1/2}.
\end{align}  
One can also add other measures such as the cost of observation to the above function.

In a two-stage decision problem, we choose an unobserved sign for observation, with the aim of maximizing the averaged objective function $\mathcal{E}(j)\equiv \langle DP(j)\rangle_O$. Note that before doing any real observation, we have access only to the probability of the outcomes $P(S_j)$; the actual or true value of an unobserved sign becomes clear only after the observation. That is why here we are taking the average over the possible outcomes, which is denoted by $\langle \cdot \rangle_O$. One can repeat the two-stage problem for $T$ times to obtain a sequence of $T$ observations: each time an optimal sign is chosen for observation followed by a real observation, which reveals the true value of the observed sign.  

In a multistage version of the problem, we want to find an optimal sequence of decisions $O_T=\{j_1,\cdots,j_T\}$, which maximizes the following objective functional of the observed signs: $\mathcal{E}[O_T]\equiv \sum_{t=1}^T \langle DP(j_t)\rangle_O$. Here, at each step, the "observed" sign takes a value that is sampled from the associated marginal probability $P(S_j)$. This probability depends on the model which we are working with. Note that here we are interested in finding an optimal sequence of observations at the beginning of the process before doing any real observation. In other words, in such a multistage problem, we are doing an "extrapolation" or "simulation" of the observation process without performing any real observation. In practice, however, we may fix the sequence of observations by a decimation algorithm: i.e., we repeat the multistage problem for $T$ times, where each time we observe the first sign suggested by the output sequence, and reduce the number of to-be-observed sings by one.     

In the following, we consider only simple approximations of the multistage problem; first we reduce the difficult multistage problem to simpler two-stage problems. More precisely, at each time step, we choose an unobserved sign $j_t$, which results to the largest disease polarization $\langle DP(j_t) \rangle_O$, for observation (greedy strategy). Then, we consider two cases: (I) we perform a real observation to reveal the true value of the suggested sign for observation, (II) we treat the suggested sign variable for observation as a stochastic variable with values that are sampled from the associated marginal probability. 

\begin{figure}
\includegraphics[width=8cm]{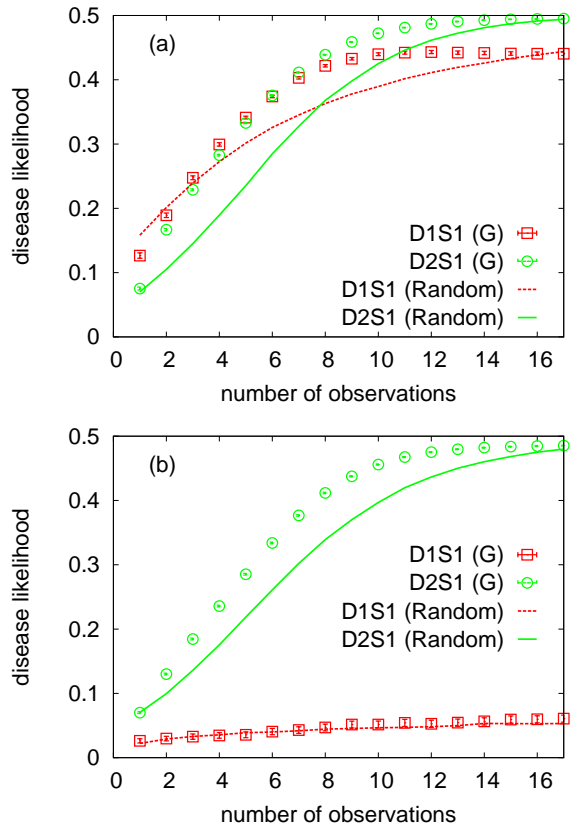} 
\caption{Overlap of the inferred disease marginals with the true hypothesis for the exponential benchmark model. The data are for the cases in which only one (a) or two (b) diseases are present. The model parameters are obtained from the conditional marginals of the true model. There are $N_D=5$ diseases, $N_S=20$ signs, and the algorithm starts with $N_O=3$ observed signs for a randomly selected hypothesis $\mathbf{D}$. An unobserved sign is chosen for observation by the greedy (G) or random strategy using the inferred probabilities, and the observed sign takes the true value given by $\mathbf{S}^*(\mathbf{D})$. 
The data are results of averaging over $1000$ independent realizations of the true model and the observation process.}\label{Q1}
\end{figure}

Let us start from the case in which we observe the true values of the signs chosen for observation. Once again, we take the exponential benchmark model given by Eq. \ref{exp-model} as the true model. We use the conditional marginals extracted from this true model to construct the simple one-disease-one-sign ($D1S1$) and two-disease-one-sign ($D2S1$) models. Suppose that we are given a disease hypothesis $\mathbf{D}$ and the associated symptoms $\mathbf{S}^*(\mathbf{D})$. We start from a few randomly chosen observed sings from the set of symptoms. Then, at each time step $t$, we compute the inferred sign probabilities $P(S_j)$, and use the greedy strategy to choose an unobserved sign for observation. The observed sign at each step takes the true value given by $\mathbf{S}^*(\mathbf{D})$. To see how much the disease probabilities obtained from the models are correlated with the true hypothesis $\mathbf{D}$, after each observation, we compute the following overlap function (or "disease likelihood"):  
\begin{align}\label{DL}
DL(t)\equiv \frac{1}{N_D}\sum_{a} (2D_a-1)\left(P(D_a=1)-\frac{1}{2}\right).
\end{align}
Besides the magnitude, our decisions also affect the way that the above quantity behaves with the number of observations.

In Fig. \ref{Q1} we see how the above overlap function, $DL(t)$, behaves for cases in which only one or two diseases are present in $\mathbf{D}$ (see also Appendix \ref{app-3}). For comparison, we also show the results obtained by a random strategy, where an unobserved sign is chosen for observation randomly and uniformly from the subset of unobserved signs. The number of sign/disease variables here is small enough to allow for an exact computation of all the marginal probabilities. It is seen that both the $D1S1$ and $D2S1$ models work very well when only one disease is present and all the other diseases are absent. The $D1S1$ model already fails when two diseases are present in the hypothesis, whereas the other model can still find the right diseases. However, we observe that even the $D2S1$ model gets confused when there are more than two diseases in the hypothesis; in such cases, we would need to consider more sophisticated models with interactions involving more than two diseases. Moreover, we observe that the difference in the performances of the greedy and random strategies decreases as the number of involved diseases increases. In Appendix \ref{app-3}, we observe similar behaviors for a more complex benchmark model $P_{true}(\mathbf{S}|\mathbf{D}) \propto 1/(1+H(\mathbf{S},\mathbf{S}^*(\mathbf{D})))$, including also the sign-sign interactions.

\begin{figure}
\includegraphics[width=14cm]{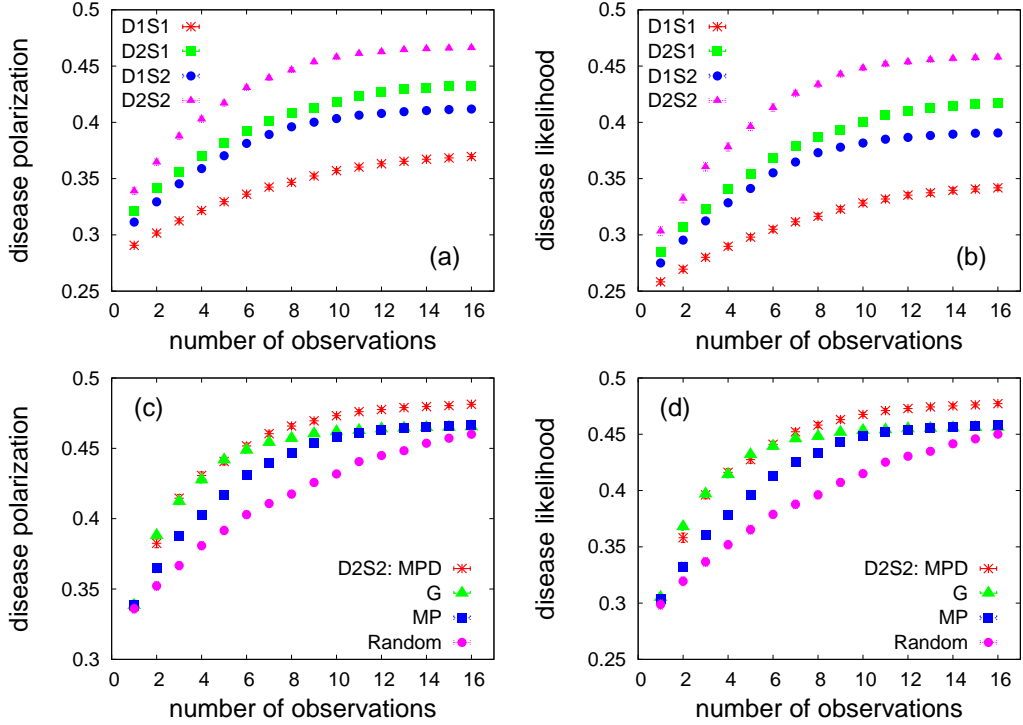} 
\caption{Diagnostic performance of the models vs the number of observations for a small number of sign/disease variables. (top) The exact disease-polarization (a) and disease-likelihood (b) obtained by the MP strategy in the one-disease-one-sign ($D1S1$), two-disease-one-sign ($D2S1$), one-disease-two-sign ($D1S2$), and two-disease-two-sign ($D2S2$) models. There are  $N_D=5$ diseases, $N_S=20$ signs, and the algorithm starts with $N_O=4$ observed signs with positive values. 
The couplings in the interaction factors are iid random numbers distributed uniformly in the specified intervals: $K_i^0=-1, K_i^{a,ab} \in (-1,+1), K_{ij}^{a,ab} \in (-1,+1)/\sqrt{N_S}$. (bottom) Comparing the exact polarization (c) and likelihood (d) of the diseases obtained by the MP, greedy (G), MPD, and random strategies, for the $D2S2$ model. The data are results of averaging over $1000$ independent realizations of the model parameters.}\label{Dt}
\end{figure}

Next, we consider the case of simulating the diagnosis process without doing any real observation. Here, we assume that an observed sign takes a value which is sampled from the associated marginal probability $P(S_j)$ at that time step. For comparison with the greedy strategy, we also introduce two other strategies for choosing an unobserved sign for observation. A naive strategy is to choose the most positive (MP) sign $j_{max}=\arg \max_j \{P(S_j=+1)\}$ for observation (MP strategy); $j_{max}$ is the sign with the maximum probability of being positive. In the early stages of the diagnosis, this probability is probably close to zero for most of the signs. So, it makes sense to choose the most positive sign for observation to obtain more information about the diseases. A more complicated strategy works by first computing the conditional probabilities $P(S_j|\mathbf{D}_{ML})$ for the maximum likelihood hypothesis $\mathbf{D}_{ML}$, and then selecting the most positive sign for observation (MPD strategy).     

To have a more general comparison of the constructed models, in the following, we assume that the model parameters ($K_i^{a,ab}$, and $K_{ij}^{a,ab}$) are iid random numbers uniformly distributed in an appropriate interval of real numbers. The leaky couplings are set to $K_i^0=-1$, which correspond to small sign probabilities $P(S_i=1|\mathrm{nodisease}) \simeq 0.05$. We assume that all the possible one-disease and two-disease interaction factors are present in the models. Moreover, inside each factor we have all the possible two-sign interactions in addition to the local sign fields. As before, the prior disease probabilities $P_0(D_a)$ are uniform probability distributions. Figure \ref{Dt} shows the performances for different models and strategies with a small number of sign/disease variables. Here, the "disease likelihood" gives the overlap of the disease probabilities with the maximum likelihood hypothesis $\mathbf{D}_{ML}$ of the models. Moreover, all the quantities are computed exactly. We see that in this case the average performance of the greedy strategy is close to that of the MPD strategy at the beginning of the process. For larger number of observations, the greedy performance degrades and approaches that of the MP strategy.

The models with disease-disease and sign-sign interactions exhibit larger polarizations of the disease probabilities and larger overlaps of the disease probabilities with the maximum-likelihood hypothesis (see also Appendix \ref{app-3}); we find that already the $D2S1$ model works considerably better than the $D1S1$ model for disease-disease interactions of relative strengths $|K_i^{ab}/K_i^a| \simeq 0.3$. A larger polarization means that we need a smaller number of observations (medical tests) to obtain more definitive disease probabilities. A larger disease likelihood, here means that we are following the direction that is suggested by the most likely diseases. In this sense, it appears that the two-sign/disease interactions could be very helpful in the early stages of the diagnosis.

We checked that the above picture also holds if we start with different numbers of observed signs, and if we double the magnitude of all the couplings. Similar behaviors are observed also for larger problem sizes (see Appendix \ref{app-3}). However, we see that for strongly interacting models with much larger higher-order interactions, e.g., $K_i^0=-1$, and $K_i^{a,ab} \in (-2,+2), K_{ij}^{a,ab} \in (-2,+2)/\sqrt{N_S}$, the MP strategy gets closer to the random strategy. The fact that the MP strategy does not work well in this strongly correlated regime was indeed expected. Here, the greedy strategy is working better than the MPD strategy, and both are still performing better than the random strategy.

\section{Discussions}\label{S3}

In summary, we showed that considering the sign-sign and disease-disease interactions can significantly change the statistical importance of the signs and diseases. More importantly, we found that these higher-order correlations could be very helpful in the process of diagnosis, especially in the early stages of the diagnosis. The results in Figs. \ref{Q1} and \ref{Dt} (and similar figures in appendices) also indicate the significance and relevance of optimization of the diagnosis procedure, where a good strategy could considerably increase the polarization and likelihood of the diseases compared to the random strategy. In addition, we devised an approximate inference algorithm with a time complexity that grows linearly with the number of interaction factors connecting the diseases to the signs, and exponentially with the maximum number of signs that are associated with such interaction factors. For clarity, in this work, we considered only algorithms of minimal structure and complexity. It would be straightforward to employ more accurate learning and inference algorithms in constructing the models and inferring the statistical information from the models.  The challenge is, of course, to go beyond the heuristic and greedy algorithms that we used to study the multistage stochastic optimization problem of deciding on the relevance and order of the medical observations.      

In this study, we considered very simple structures for the prior probability of the diseases $P_0(\mathbf{D})$ and the leak probability of the signs $P(\mathbf{S}|\mathrm{nodisease})$. Obviously, depending on the available statistical data, we can obtain  more reliable models also for these probability distributions. Alternatively, we could employ the maximum entropy principle to construct directly the joint probability distribution of the sign and disease variables using the joint marginal probabilities $P(S_i,S_j;D_a,D_b)$. Note that, in practice, it is easier to obtain this type of information than the stronger conditional probabilities $P(S_i,S_j|\mathrm{only} D_a)$ and $P(S_i,S_j|\mathrm{only} D_a,D_b)$. However, these measures are easier to model (or estimated by experts), in the absence of enough observational data, because they present the sole effects of single (or few) diseases.     

The emphasize, in this study, was more on the diagnostic performances of the inferred models than on the statistical significance of the selected models for a given set of clinical data. A more accurate treatment of the model selection, for a finite collection of data, accounts also the complexity of the models to avoid the over-fitting of the data. However, we note that including the sign-sign and disease-disease interactions in the models is indeed more natural than ignoring such correlations. Finally, to take into account the role of noises in the model parameters, one should take the average of the objective function over the probability distribution of the parameters, which is provided by the likelihood of the model parameters.     

Our proposed framework can be adapted to address assignment problems in cell biology, immunology, and evolutionary biology \cite{Barabasi2004, Kim2016, Ebrahim2016,Candia2013}. In contrast to clinical problems, here data availability might be less of a problem in near future. Advances in genomics, transcriptomics, proteomics and metabolomics promise high resolution molecular characterization of cells. Intensive research has also been directed towards live single cell analysis which allows characterization of pathways from an initial state to a final state \cite{spiller2010}. Our approach can be used to do early assignments and thus not only provides accuracy but also an improved sensitivity for diagnostics at the cellular level.      


\appendix

\section{Details of the models}\label{app-1}
In this section, we give more details of the models, and show how the model parameters are computed given the following true marginal probabilities:
\begin{align}
P_{true}(S_i|\mathrm{nodisease}),&\hskip1cm \forall i,\\
P_{true}(S_i,S_j|\mathrm{only} D_a),&\hskip1cm \forall i\neq j \& a,\\
P_{true}(S_i,S_j|\mathrm{only} D_a,D_b),&\hskip1cm \forall i \neq j \& a\neq b.
\end{align}

\subsection{Case of one-disease-one-sign interactions: $D1S1$ model}\label{app-11}
The simplest model is obtained by considering only the one-disease factors besides the leaky interactions.
In addition, we assume that each factor contains only local sign fields, i.e., 
\begin{align}
\phi_0(\mathbf{S}) &= e^{\sum_i K_i^0 S_i},\\
\phi_{a}(\mathbf{S}|D_a) &= e^{D_{a}\sum_i K_i^{a} S_i},\hskip1cm
\phi_{ab}(\mathbf{S}|D_a,D_b) =\cdots=1.
\end{align}  
Here, the couplings are 
\begin{align}
K_i^0 &=\frac{1}{2}\ln\left(\frac{P_{true}(S_i=+1|\mathrm{nodisease})}{P_{true}(S_i=-1|\mathrm{nodisease})}\right),\\
K_i^{a} &=\frac{1}{2}\ln\left(\frac{P_{true}(S_i=+1|\mathrm{only} D_a)}{P_{true}(S_i=-1|\mathrm{only} D_a)}\right)-K_i^0.
\end{align}  

Then the partition function reads
\begin{align}
Z(\mathbf{D})=\prod_i \left( 2\cosh(K_i^0+\sum_{a}D_{a}K_i^{a}) \right).
\end{align}  
The likelihood of diseases is given by
\begin{align}
\mathcal{L}(\mathbf{D}|\mathbf{S}^o)=\prod_{i\in O} \left(\frac{e^{(K_i^0+\sum_{a}D_{a}K_i^{a})S_i}}{2\cosh(K_i^0+\sum_{a}D_{a}K_i^{a})} \right)
\times \prod_{a} P_{0}(D_{a}).
\end{align}  
Note that after summing over the sign variables the disease variables become correlated and the problem of computing the marginals 
is not straightforward. The same problem arises in computation of the probability distribution of unobserved signs,
\begin{align}
\mathcal{M}(\mathbf{S}^u|\mathbf{S}^o) &=\sum_{\mathbf{D}}\prod_{i} \left(\frac{e^{(K_i^0+\sum_{a}D_{a}K_i^{a})S_i}}{2\cosh(K_i^0+\sum_{a}D_{a}K_i^{a})} \right)
\times \prod_{a} P_{0}(D_{a}). 
\end{align}  
For the moment, we assume the number of involved variables $N_{D,S}$ are small enough to do the above computations exactly. For larger number of variables, we have to resort to an approximate algorithm. We will describe such an approximate algorithm later when we consider more serious examples.

\subsection{Case of two-disease-one-sign interactions: $D2S1$ model}\label{app-12}
Now we consider the more interesting case of one- and two-disease interaction factors.
Again, we assume that each factor contains only local sign fields, i.e., 
\begin{align}
\phi_0(\mathbf{S}) &= e^{\sum_i K_i^0 S_i},\\
\phi_{a}(\mathbf{S}|D_a) &= e^{D_{a}\sum_i K_i^{a} S_i},\\
\phi_{ab}(\mathbf{S}|D_a,D_b) &= e^{D_{a}D_{b}\sum_i K_i^{ab} S_i},\hskip1cm
\phi_{abc}(\mathbf{S}|D_a,D_b,D_c)=\cdots=1.
\end{align}
Here, the couplings are 
\begin{align}
K_i^0 &=\frac{1}{2}\ln\left(\frac{P_{true}(S_i=+1|\mathrm{nodisease})}{P_{true}(S_i=-1|\mathrm{nodisease})}\right),\\
K_i^{a} &=\frac{1}{2}\ln\left(\frac{P_{true}(S_i=+1|\mathrm{only} D_a)}{P_{true}(S_i=-1|\mathrm{only} D_a)}\right)-K_i^0,\\
K_i^{ab} &=\frac{1}{2}\ln\left(\frac{P_{true}(S_i=+1|\mathrm{only} D_a,D_b)}{P_{true}(S_i=-1|\mathrm{only} D_a,D_b)}\right)-K_i^0-K_i^a-K_i^b.
\end{align}  
  
Still, we can easily obtain the partition function  
\begin{align}
Z(\mathbf{D})=\prod_i \left( 2\cosh(K_i^0+\sum_{a}D_{a}K_i^{a}+\sum_{a<b}D_{a}D_{b}K_i^{ab}) \right).
\end{align}  
The likelihood reads  
\begin{align}
\mathcal{L}(\mathbf{D}|\mathbf{S}^o) &=\prod_{i\in O} \left(\frac{e^{(K_i^0+\sum_{a}D_{a}K_i^{a}+\sum_{a<b}D_{a}D_{b}K_i^{ab})S_i}}{2\cosh(K_i^0+\sum_{a}D_{a}K_i^{a}+\sum_{a<b}D_{a}D_{b}K_i^{ab})} \right)
\times \prod_{a} P_{0}(D_{a}).
\end{align}  
Similarly,
\begin{align}
\mathcal{M}(\mathbf{S}^u|\mathbf{S}^o) &=\sum_{\mathbf{D}}\prod_{i} \left(\frac{e^{(K_i^0+\sum_{a}D_{a}K_i^{a}+\sum_{a<b}D_{a}D_{b}K_i^{ab})S_i}}{2\cosh(K_i^0+\sum_{a}D_{a}K_i^{a}+\sum_{a<b}D_{a}D_{b}K_i^{ab})} \right)
\times \prod_{a} P_{0}(D_{a}).
\end{align}

\subsection{Introducing the two-sign interactions: $D1S2$ and $D2S2$ models}\label{app-13}
A more challenging model is obtained with the two-sign interactions  
\begin{align}
\phi_0(\mathbf{S}) &= e^{\sum_i K_i^0 S_i},\\
\phi_{a}(\mathbf{S}|D_a) &= e^{D_{a}[\sum_i K_i^{a} S_i+\sum_{i<j} K_{ij}^{a} S_iS_j]},\\
\phi_{ab}(\mathbf{S}|D_a,D_b) &= e^{D_{a}D_{b}[\sum_i K_i^{ab} S_i+\sum_{i<j} K_{ij}^{ab} S_iS_j]},\hskip1cm
\phi_{abc}(\mathbf{S}|D_a,D_b,D_c)=\cdots=1.
\end{align}  
Here, even computing the partition function is difficult, and from the beginning we have to resort to approximations.  

First of all, we need to obtain the couplings given the conditional probabilities,
\begin{align}
P_0(S_i) &\equiv P_{true}(S_i|\mathrm{nodisease}),\\ 
P_a(S_i,S_j) &\equiv P_{true}(S_i,S_j|\mathrm{only} D_a),\hskip0.5cm P_a(S_i) \equiv P_{true}(S_i|\mathrm{only} D_a),\\
P_{ab}(S_i,S_j) &\equiv P_{true}(S_i,S_j|\mathrm{only} D_a,D_b), \hskip0.5cm P_{ab}(S_i) \equiv P_{true}(S_i|\mathrm{only} D_a,D_b).
\end{align}  
As before, the zero-order couplings are given by
\begin{align}
K_i^0 =\frac{1}{2}\ln\left(\frac{P_0(S_i=+1)}{P_0(S_i=-1)}\right).
\end{align}  
To obtain the other couplings, we resort to the so-called independent-pair approximation \cite{RAH-cn-2009}, where we assume
\begin{align}
P_a(S_i,S_j)&\propto e^{K_{ij}^aS_iS_j+h_{ij}^aS_i+h_{ji}^aS_j},\hskip0.5cm P_a(S_i)\propto e^{h_{i}^aS_i}\\
P_{ab}(S_i,S_j)&\propto e^{K_{ij}^{ab}S_iS_j+h_{ij}^{ab}S_i+h_{ji}^{ab}S_j},\hskip0.5cm P_{ab}(S_i)\propto e^{h_{i}^{ab}S_i}.
\end{align}  
In this way we obtain,
\begin{align}
K_{ij}^a &=\frac{1}{4}\ln\left(\frac{P_a(S_i=+1,S_j=+1)P_a(S_i=-1,S_j=-1)}{P_a(S_i=+1,S_j=-1)P_a(S_i=-1,S_j=+1)}\right),\\
K_{ij}^{ab} &=\frac{1}{4}\ln\left(\frac{P_{ab}(S_i=+1,S_j=+1)P_{ab}(S_i=-1,S_j=-1)}{P_{ab}(S_i=+1,S_j=-1)P_{ab}(S_i=-1,S_j=+1)}\right)-K_{ij}^a-K_{ij}^b,
\end{align}  
and
\begin{multline}
K_{i}^a =\frac{1}{4}\sum_{j\neq i}\ln\left(\frac{P_a(S_i=+1,S_j=+1)P_a(S_i=+1,S_j=-1)}{P_a(S_i=-1,S_j=-1)P_a(S_i=-1,S_j=+1)}\right)\\
-\frac{(N_S-2)}{2}\ln\left(\frac{P_a(S_i=+1)}{P_a(S_i=-1)}\right)-K_i^0,
\end{multline}
\begin{multline}
K_{i}^{ab} =\frac{1}{4}\sum_{j\neq i}\ln\left(\frac{P_{ab}(S_i=+1,S_j=+1)P_{ab}(S_i=+1,S_j=-1)}{P_{ab}(S_i=-1,S_j=-1)P_{ab}(S_i=-1,S_j=+1)}\right)\\
-\frac{(N_S-2)}{2}\ln\left(\frac{P_{ab}(S_i=+1)}{P_{ab}(S_i=-1)}\right)-K_i^0-K_i^a-K_i^b.
\end{multline}  
One can find other approximation methods in the literature to address the above inverse problem. For example, one can employ the exact fluctuation-response relations within the Bethe approximation to obtain more accurate estimations for the marginal probability distribution of sign variables \cite{NB-prl-2012,T-jstat-2012}.  

\begin{figure}
\includegraphics[width=8cm]{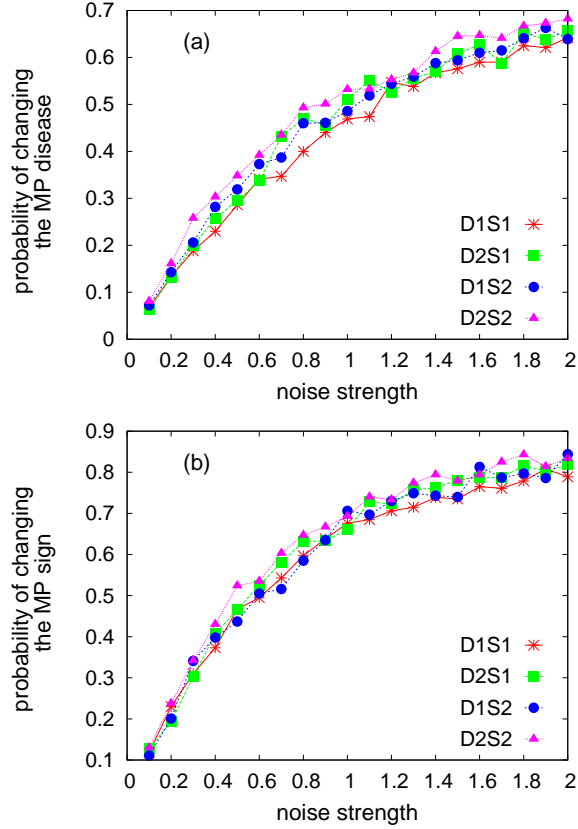} 
\caption{Probability of changing the most probable (MP) disease (a) and sign (b) variable in the one-disease-one-sign ($D1S1$), two-disease-one-sign ($D2S1$), one-disease-two-sign ($D1S2$), and two-disease-two-sign ($D2S2$) models. There are  $N_D=5$ diseases, $N_S=15$ signs, and $N_O=3$ observed signs with positive values.  The couplings in the interaction factors are iid random numbers distributed uniformly in the specified intervals: $K_i^0=-1, K_i^{a,ab} \in (-1,+1), K_{ij}^{a,ab} \in (-1,+1)/\sqrt{N_S}$. A noise of strength $\epsilon$ changes the couplings by $\delta K_i^{a,ab} \in  (-\epsilon,+\epsilon), \delta K_{ij}^{a,ab} \in (-\epsilon,+\epsilon)/\sqrt{N_S}$. The data have been obtained from $1000$ independent realizations of the model parameters.}\label{noise}
\end{figure}

Note that the model parameters are usually noisy because they may come from an approximate learning algorithm with a finite collection of clinical data. Thus, it is important to know how the noise strength affects the marginal sign/disease probabilities. In Fig. \ref{noise}, we show how much the uncertainties in the model parameters are likely to change the sign/disease with the highest probability of being positive/present. We see that this probability increases linearly with the relative strength of the noise in the model parameters $\epsilon\equiv \delta K/K$; for example, in a small model of $N_D=5$ diseases, $N_S=15$ signs with $N_O=3$ observed signs, a noise of strength $\epsilon=0.1$ changes the most probable sign/disease in nearly $\% 10$ of the cases.

\section{An approximate inference algorithm}\label{app-2}
In this section, we present an approximate way of computing the sign/disease marginal probabilities. To be specific, we will focus on the two-disease-two-sign ($D2S2$) model.   

We will consider the joint measure 
\begin{align}
P(\mathbf{S};\mathbf{D})= P(\mathbf{S}|\mathbf{D})P_0(\mathbf{D}) \propto e^{-\mathcal{H}(\mathbf{S};\mathbf{D})},
\end{align}  
of the disease and sign variables. Here, we defined the energy function
\begin{multline}
\mathcal{H}(\mathbf{S};\mathbf{D}) \equiv -\sum_{i} K_i^0 S_i - \sum_{a}D_{a}[\sum_{i} K_i^{a} S_i+\sum_{i<j} K_{ij}^{a} S_iS_j]-\\
\sum_{a<b}D_{a}D_{b}[\sum_{i} K_i^{ab} S_i+\sum_{i<j} K_{ij}^{ab} S_iS_j]+\ln Z(\mathbf{D})-\ln P_0(\mathbf{D}).
\end{multline}
We may obtain the local probability marginals by a Monte Carlo (MC) algorithm for the above system of interacting variables.   
But computing the above energy function for a given configuration of the variables is computationally hard because computing $Z(\mathbf{D})$ is in general hard. Thus, to reduce the computation time we have to find a good approximation for the partition function.     
 
The exact expression for the partition function is
\begin{align}
Z(\mathbf{D})=\sum_{\mathbf{S}} \phi_0(\mathbf{S}) \times \prod_{a} \phi_{a}(\mathbf{S}|D_a)\times  \prod_{a < b} \phi_{ab}(\mathbf{S}|D_a,D_b), 
\end{align}  
which in terms of the couplings reads
\begin{align}
Z(\mathbf{D}) &=\sum_{\mathbf{S}} e^{\sum_i h_i(\mathbf{D}) S_i+\sum_{i<j}J_{ij}(\mathbf{D})S_iS_j},\\
h_i(\mathbf{D}) &\equiv K_i^0+\sum_{a}D_{a}K_i^{a}+\sum_{a<b}D_{a}D_{b}K_i^{ab},\\
J_{ij}(\mathbf{D}) &\equiv \sum_{a}D_{a}K_{ij}^{a}+\sum_{a<b}D_{a}D_{b}K_{ij}^{ab}.
\end{align}  
We are going to use the Bethe approximation to find an estimation for the above partition function \cite{Pearl,KFL-inform-2001,MM-book-2009}. More precisely,
in this approximation, we assume that the interaction graph of the variables (defined by the couplings $J_{ij}$) is a tree graph.
We know how to compute exactly and efficiently the local marginal probabilities and the partition function in tree interaction graphs. The fact that there is no loop in a tree graph allows us to write the marginal probability of a variable in terms of the cavity probabilities of the neighboring variables, computed in the absence of the interested variable. These cavity probabilities are the main quantities of the Bethe approximation. Here, e.g., we need to compute the cavity marginals $\mu_{i\to j}(S_i)$, giving the probability of value $S_i$ for sign $i$ in the absence of sign $j$. The Bethe equations governing these cavity marginals
are the following recursive equations \cite{MM-book-2009}:
\begin{align}
\mu_{i\to j}(S_i)=\frac{1}{z_{i\to j}}e^{h_iS_i}\prod_{k\neq i,j}\left( \sum_{S_k} e^{J_{ik}S_iS_k} \mu_{k \to i}(S_k)\right),
\end{align}  
where $z_{i\to j}$ is a normalization constant.
We can solve these equations by iteration starting from random initial cavity marginals; the cavity marginals are updated according to the above equations in a random and sequential way, until the algorithm converges. Then, we use these stationary cavity marginals to compute the local probabilities and the partition function \cite{MM-book-2009}.  

In this way, for the partition function within the Bethe approximation, we obtain
\begin{align}
Z(\mathbf{D}) &\simeq e^{\sum_i\Delta_i(\mathbf{D})-\sum_{i<j}\Delta_{ij}(\mathbf{D})}, \label{ZD}\\
e^{\Delta_i(\mathbf{D})} &\equiv \sum_{S_i} e^{h_iS_i}\prod_{j\neq i}\left( \sum_{S_j} e^{J_{ij}S_iS_j} \mu_{j \to i}(S_j)\right),\label{di}\\
e^{\Delta_{ij}(\mathbf{D})} &\equiv \sum_{S_i,S_j} e^{J_{ij}S_iS_j} \mu_{i \to j}(S_i)\mu_{j \to i}(S_j).\label{dij} 
\end{align}  

Notice that here the cavity marginals $\mu_{i\to j}(S_i)$ depend implicitly on the disease variables. Now, we can compute efficiently an approximate change in the energy function $\mathcal{H}(\mathbf{S};\mathbf{D})$ when a disease variable is chosen for updating in a Monte Carlo algorithm. In the same way, we may find an estimate of the likelihood of diseases by
\begin{align}
\mathcal{L}(\mathbf{D}|\mathbf{S}^o)=\frac{Z(\mathbf{D}|\mathbf{S}^o)}{Z(\mathbf{D})}P_0(\mathbf{D}),
\end{align}  
where
\begin{align}
Z(\mathbf{D}|\mathbf{S}^o) =\sum_{\mathbf{S}^u} \phi_0(\mathbf{S}) \times \prod_{a} \phi_{a}(\mathbf{S}|D_a)\times  \prod_{a < b} \phi_{ab}(\mathbf{S}|D_a,D_b), 
\end{align}  
can be estimated within the Bethe approximation, as we did for $Z(\mathbf{D})$, but with fixed values for the observed signs.
This approximate expression for the likelihood is useful to find the most likely diseases $\mathbf{D}_{ML}$ maximizing the likelihood.   

Finding the maximum likelihood hypothesis $\mathbf{D}_{ML}$, and obtaining good estimations for the probability marginals of $P(\mathbf{S};\mathbf{D})$ could be still very time consuming for a large number of variables and frustrating energy functions. In the following, we will utilize the Bethe approximation to estimate the above marginal probabilities.

\subsection{Belief propagation equations}\label{app-21} 
The interaction factors associated to diseases $a$ and $(ab)$ are
\begin{align}
\phi_a &= \exp\left( D_{a}[\sum_i K_{i}^{a}S_i+\sum_{i<j}K_{ij}^{a}S_iS_j] \right),\\
\phi_{ab} &= \exp\left( D_{a}D_{b}[\sum_iK_{i}^{ab}S_i+\sum_{i<j}K_{ij}^{ab}S_iS_j] \right),
\end{align} 
respectively. Let us label the above interactions with $\alpha \in \{a,b,\dots,(ab),(ac),\dots\}$ and use $\phi_{\alpha}(\mathbf{S}^{\alpha}|\mathbf{D}^{\alpha})$ for the corresponding interaction factors. We represent the subset of sign and disease variables in an interaction factor with $\partial_S \alpha$ and $ \partial_D \alpha$. The subset of interaction factors depending on sign $i$ and disease $a$ are denoted by $\partial i$ and $\partial a$, respectively. We will assume that the resulting interaction graph is sparse, that is the numbers of interaction factors $(M_a,M_{ab})$ are small compared to $N_S$, and the numbers of sign variables involved in the interaction factors $(k_a,k_{ab})$ are of order one. We take $\nu_{i\to \alpha}(S_i)$ for the probability of having value $S_i$ for sign $i$ in the absence of interaction factor $\alpha$. Similarly, define the cavity probabilities $\nu_{a\to \alpha}(D_a)$. In the Bethe approximation, we write approximate expressions for these cavity marginals in terms of the messages received from the other interaction factors in the system \cite{KFL-inform-2001,MM-book-2009}. More precisely,
\begin{align}
\nu_{i\to \alpha}(S_i) &\propto e^{K_i^0S_i}\prod_{\beta \in \partial i\neq \alpha} \Psi_{\beta \to i}(S_i),\\
\nu_{a \to \alpha}(D_a) &\propto P_{0}(D_{a})\prod_{\beta \in \partial a\neq \alpha} \Psi_{\beta \to a}(D_a) \Psi_{Z\to a}(D_{a}).
\end{align} 
Note that the equation for $\nu_{i\to \alpha}(S_i)$ gives the cavity marginal for an unobserved sign; the cavity marginals of observed signs are fixed to values determined by the observations.
The cavity messages received from the interaction factors are given by
\begin{align}
\Psi_{\alpha \to i}(S_i) \propto \sum_{\{D_a:a \in \partial_D \alpha\}}\sum_{\{S_j:j\in \partial_S \alpha\setminus i \}} \phi_{\alpha}(\mathbf{S}^{\alpha}|\mathbf{D}^{\alpha})\prod_{a\in \partial_D \alpha}\nu_{a\to \alpha}(D_a)\prod_{j\in \partial_S \alpha\setminus i}\nu_{j\to \alpha}(S_j),\\
\Psi_{\alpha \to a}(D_a) \propto \sum_{\{D_b:b \in \partial_D \alpha\setminus a\}}\sum_{\{S_j:j\in \partial_S \alpha\}} \phi_{\alpha}(\mathbf{S}^{\alpha}|\mathbf{D}^{\alpha})\prod_{b\in \partial_D \alpha\setminus a}\nu_{b\to \alpha}(D_b)\prod_{j\in \partial_S \alpha}\nu_{j\to \alpha}(S_j).
\end{align}

It remains to write the cavity marginals related to the partition function $Z(\mathbf{D})$,    
\begin{align}
\nu_{a \to Z}(D_a) &\propto P_{0}(D_a) \prod_{\alpha \in \partial a} \Psi_{\alpha\to a}(D_a),\label{MZ}\\ 
\Psi_{Z\to a}(D_a) &\propto \sum_{\{D_b:b \neq a\}} e^{-\ln Z(\mathbf{D})} \prod_{b \neq a}\nu_{b\to Z}(D_b).\label{PSIZ}
\end{align} 
The main difficulty comes from the last equation, where one has to sum over all possible values of the disease variables and compute $Z(\mathbf{D})$ for each configuration. We already know how to compute the partition function by the Bethe approximation. But, the sum over the disease variables grows exponentially with the number of variables. A naive approximation to get around this complexity is obtained by a Bethe factorization of the partition function; this is a nonnegative function of the disease variables, therefore,  up to a normalization constant, it can be interpreted as a probability measure of the associated variables. Suppose $\mu_a(D_a)$ and $\mu_{\alpha}(\mathbf{D}^{\alpha})$ are the local probability marginals of this measure. Then we may approximate the partition function (up to a constant multiplication factor) by 
\begin{align}
Z(\mathbf{D}) \propto \prod_a \mu_a(D_a)\prod_{\alpha}\left(\frac{\mu_{\alpha}(\mathbf{D}^{\alpha})}{\prod_{a \in \partial_D \alpha}\mu_a(D_a)} \right).
\end{align} 
Note that, we can obtain approximate expressions for the above local marginals by solving the above equations without considering the interactions induced by the partition function $Z(\mathbf{D})$.       
Moreover, the above approximation can be improved by considering the higher-order marginals in the above factorization.  

\begin{figure}
\includegraphics[width=8cm]{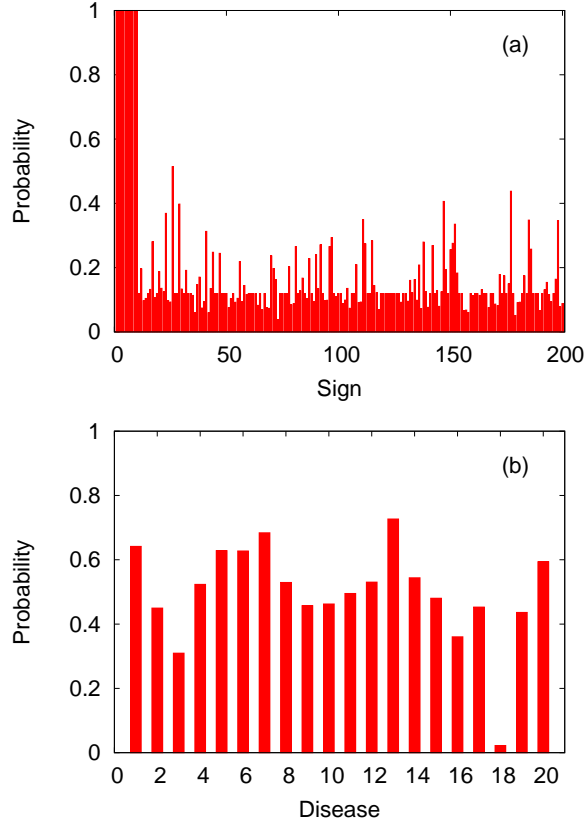} 
\caption{The sign (a) and disease (b) marginal probabilities computed by the approximate belief propagation algorithm for a single instance of the two-disease-two-sign ($D2S2$) model. The parameters are $N_D=20, N_S=200, N_O=10, M_a=20, M_{ab}=40, k_{a}=6, k_{ab}=3$. Each factor includes all the possible two-sign interactions in addition to the local sign fields. The couplings are iid random numbers distributed uniformly in the specified intervals: $K_i^0=-1, K_i^{a,ab} \in (-1,+1)$, and $K_{ij}^{a} \in (-1,+1)/\sqrt{k_a}, K_{ij}^{ab} \in (-1,+1)/\sqrt{k_{ab}}$. The sign and disease variables in each factor are chosen randomly with a uniform probability.}\label{PD2S2-BP}
\end{figure}

Now that $Z(\mathbf{D})$ is factorized, we do not need any longer the cavity messages introduced in Eqs. \ref{MZ} and \ref{PSIZ}.      
Instead, we will absorb the factors in the interaction factors $\phi_{\alpha}$, working with the modified interactions 
\begin{align}
\tilde{\phi}_a &\equiv  \frac{\phi_a}{\mu_a(D_a)},\\
\tilde{\phi}_{ab} &\equiv \mu_a(D_a)\mu_b(D_b)\frac{\phi_{ab}}{\mu_{ab}(D_a,D_b)}.
\end{align} 
In summary, within the above approximation of the partition function, we first solve the following equations for the cavity marginals, considering the modified interactions,
\begin{align}
\nu_{i\to \alpha}(S_i) &\propto e^{K_i^0S_i}\prod_{\beta \in \partial i \neq \alpha} \tilde{\Psi}_{\beta \to i}(S_i),\label{Mi}\\
\nu_{a \to \alpha}(D_a) &\propto P_{0}(D_{a})\prod_{\beta \in \partial a \neq \alpha} \tilde{\Psi}_{\beta \to a}(D_a),\label{Ma}
\end{align} 
and
\begin{align}
\tilde{\Psi}_{\alpha \to i}(S_i) \propto \sum_{\{D_a:a \in \partial_D \alpha\}}\sum_{\{S_j:j\in \partial_S \alpha\setminus i \}} \tilde{\phi}_{\alpha}(\mathbf{S}^{\alpha}|\mathbf{D}^{\alpha})\prod_{a\in \partial_D \alpha}\nu_{a\to \alpha}(D_a)\prod_{j\in \partial_S \alpha\setminus i}\nu_{j\to \alpha}(S_j),\label{Psi}\\
\tilde{\Psi}_{\alpha \to a}(D_a) \propto \sum_{\{D_b:b \in \partial_D \alpha\setminus a\}}\sum_{\{S_j:j\in \partial_S \alpha\}} \tilde{\phi}_{\alpha}(\mathbf{S}^{\alpha}|\mathbf{D}^{\alpha})\prod_{b\in \partial_D \alpha\setminus a}\nu_{b\to \alpha}(D_b)\prod_{j\in \partial_S \alpha}\nu_{j\to \alpha}(S_j).\label{Psa}
\end{align}
These equations are solved by iteration; we start from random initial messages, and update the cavity marginals according to the above equations until the algorithm converges.      
Then, we obtain the sign and disease marginal probabilities from similar equations, but taking into account all the neighboring interactions, that is,
\begin{align}
P(S_i) &\propto e^{K_i^0S_i}\prod_{\alpha \in \partial i} \tilde{\Psi}_{\alpha \to i}(S_i),\\
P(D_a) &\propto P_{0}(D_{a})\prod_{\alpha \in \partial a} \tilde{\Psi}_{\alpha \to a}(D_a).
\end{align}

\begin{figure}
\includegraphics[width=8cm]{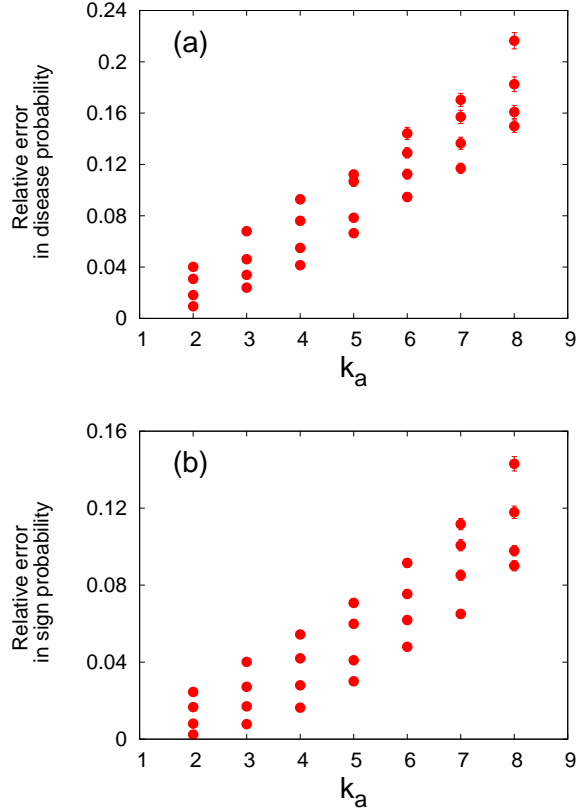} 
\caption{Relative errors of the disease (a) and sign (b) marginal probabilities computed by the approximate belief propagation algorithm for the two-disease-two-sign ($D2S2$) model. The parameters are $N_D=5, N_S=20, N_O=3, M_a=5, k_{ab}=2$, with $M_{ab}$ and $k_a$ taking different values. Each factor includes all the possible two-sign interactions in addition to the local sign fields. The couplings are iid random numbers distributed uniformly in the specified intervals: $K_i^0=-1, K_i^{a,ab} \in (-1,+1)$, and $K_{ij}^{a} \in (-1,+1)/\sqrt{k_a}, K_{ij}^{ab} \in (-1,+1)/\sqrt{k_{ab}}$. The sign and disease variables in each factor are chosen randomly with a uniform probability.}\label{eBP}
\end{figure}
  
Figure \ref{PD2S2-BP} shows the sign and disease probabilities we obtain for the two-disease-two-sign ($D2S2$) model with $N_D=20$ diseases, $N_S=200$ signs, and $N_O=10$ observed signs of positive values. Here, we take a sparse interaction graph with $M_a=20$ one-disease factors and $M_{ab}=40$ two-disease factors connected randomly to $k_a=6$ and $k_{ab}=3$ sign variables, respectively.  The quality of our approximation depends very much on the structure of the interaction factors and strength of the associated couplings in the models. The Bethe approximation is exact for interaction graphs that have a tree structure. It is expected to work very well also in sparsely connected graphs, where the number of interaction factors ($M_a, M_{ab}$) and the number of signs associated to an interaction factor ($k_a, k_{ab}$) are small compared to the total number of sign variables. This behavior is observed in Fig. \ref{eBP}, which displays the relative errors in the marginal probabilities of sign/disease variables, obtained by the above approximate algorithm. The time complexity of our approximate inference algorithm grows linearly with the number of interaction factors, and exponentially with the number of variables involved in such interactions; with $N_D=500, N_S=5000, M_a=500, M_{ab}=1000, k_a=10, k_{ab}=5$ the algorithm takes about one minute of CPU time on a standard PC to compute the local marginals.

\section{Comparing the diagnostic performance of the models}\label{app-3}
As explained in the main text, we reduce the multistage problem of diagnosis to a sequence of two-stage problems. At each step, we use a greedy or heuristic
strategy to choose the next sign for observation. We study two different cases depending on the values that are assigned to the observed signs: (I) in a real observation we find the true value of the observed sign, (II) in a simulation of the observation process, the value of observed sign $j$ is sampled from its marginal probability $P(S_j)$. In this section, we present more details of the two cases.

\subsection{Case I}\label{app-31}
Here, we need to know the true sign values, therefore, we choose a true model with a clear association of symptoms to diseases.  
As the benchmark, we take the exponential conditional probability,
\begin{align}
P_{true}(\mathbf{S}|\mathbf{D})=\frac{1}{Z_{true}(\mathbf{D})}e^{-H(\mathbf{S},\mathbf{S}^*(\mathbf{D}))},
\end{align}  
where $\mathbf{S}^*(\mathbf{D})$ gives the symptoms of hypothesis $\mathbf{D}$. We choose the symptoms randomly and uniformly from the space of sign configurations. Here $H(\mathbf{S},\mathbf{S}')=\sum_{i=1}^{N_S} (S_i-S_i')^2/4$ is the Hamming distance (number of different signs) of the two sign configurations. As before, we will take a factorized and unbiased prior disease probability $P_0(\mathbf{D})$. Note that there is no sign-sign interaction in the above true model. Therefore, we need to  consider only the effects of disease-disease interactions, and study the $D1S1$ and $D2S1$ models. Let us assume that we are given the true conditional marginals $P_{true}(S_i|\mathrm{nodisease})$, $P_{true}(S_i|\mathrm{only}D_a)$, and $P_{true}(S_i|\mathrm{only}D_a,D_b)$. From these information, we can exactly obtain the model parameters $K_i^0$, $K_i^a$, and $K_i^{ab}$ as described in Appendix \ref{app-1}.     

\begin{figure}
\includegraphics[width=8cm]{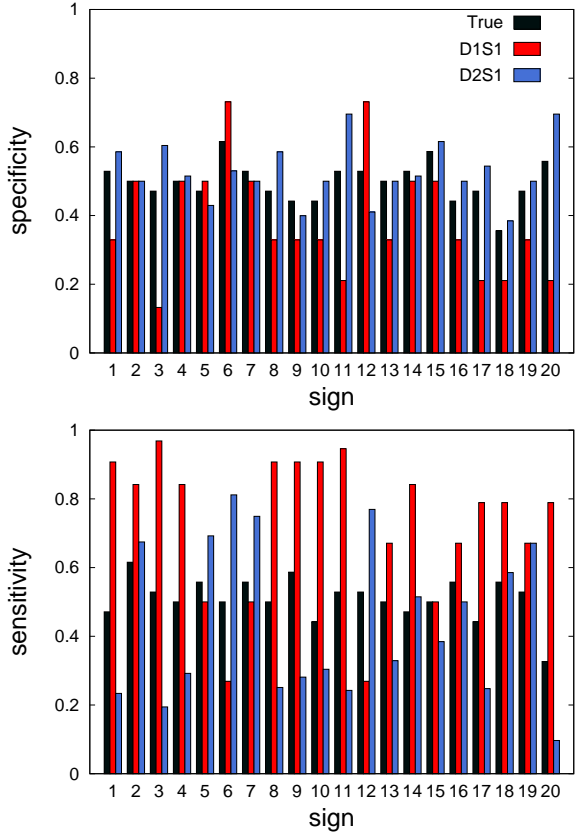} 
\caption{Sensitivity, $P(S_i=+1|D_a=1)$, and specificity, $P(S_i=-1|D_a=0)$, in the $D1S1$ and $D2S1$ models for the exponential benchmark model. The data are for one of the diseases in a single realization of the true model. The model parameters are obtained from the conditional marginals of the exponential true model. Here we have $N_D=5$ diseases, $N_S=20$ signs.}\label{Snp-2}
\end{figure}

\begin{figure}
\includegraphics[width=8cm]{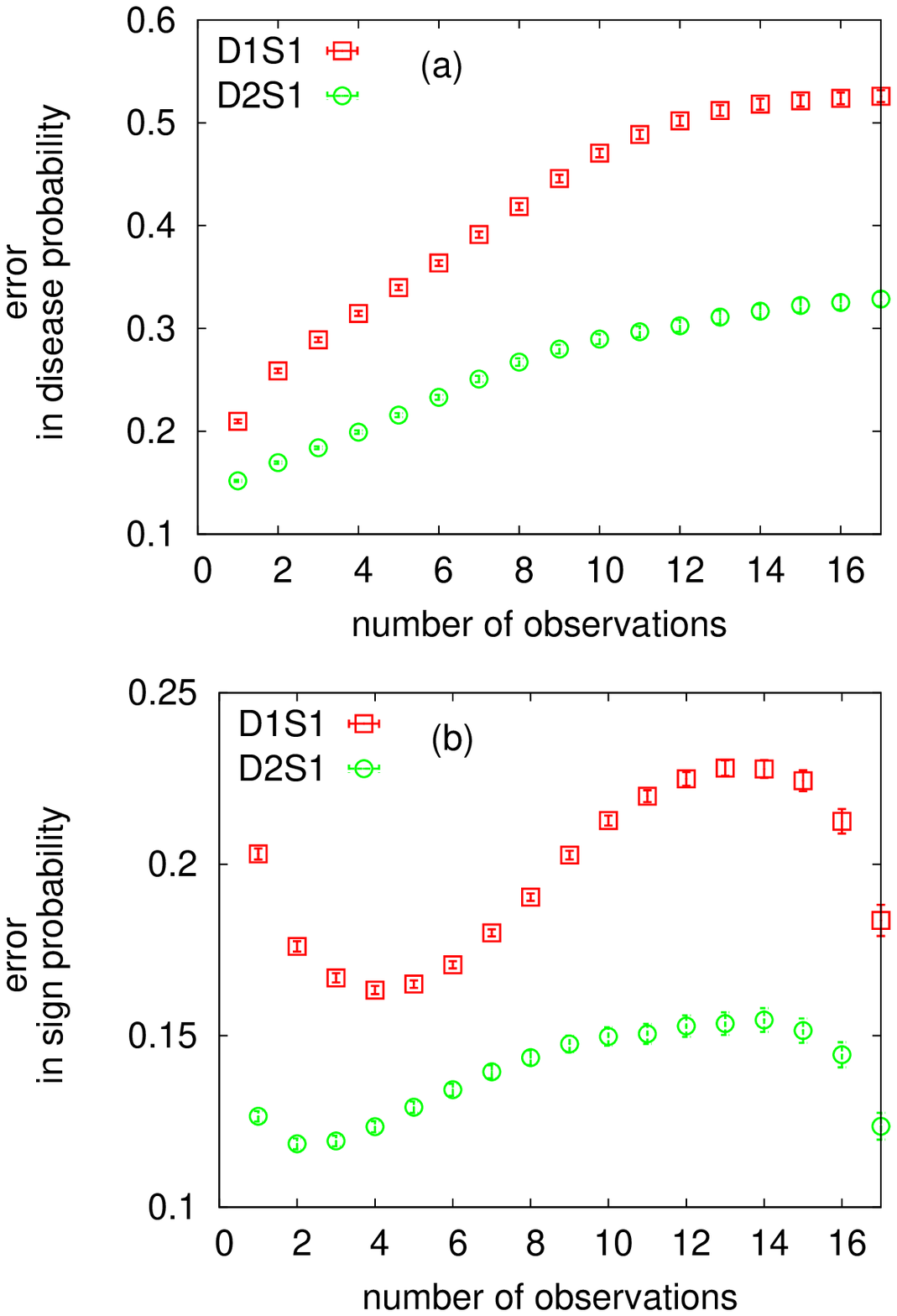} 
\caption{The RMS error in disease (a) and  sign (b) marginal probabilities in the $D1S1$ and $D2S1$ models for the exponential benchmark model. The model parameters are obtained from the conditional marginals of the true model. There are  $N_D=5$ diseases, $N_S=20$ signs, and the algorithm starts with $N_O=3$ observed signs for a randomly selected hypothesis $\mathbf{D}$. An unobserved sign is chosen for observation by the MP strategy using the inferred sign probabilities, and the observed sign takes the true value given by $\mathbf{S}^*(\mathbf{D})$. The data are results of averaging over $1000$ independent realizations of the true model and the observation process.}\label{Q0}
\end{figure}

Let us take a single realization of the true model for a small number of sign/disease variables. This allows us to compute exactly the conditional marginal probabilities $P(S_i=+1|D_a=1)$ and $P(S_i=-1|D_a=0)$, which are known in the literature as sensitivity and specificity, respectively. As Fig. \ref{Snp-2} shows, both the probabilities are in average closer to the true values in the $D2S1$ model.    

\begin{figure}
\includegraphics[width=14cm]{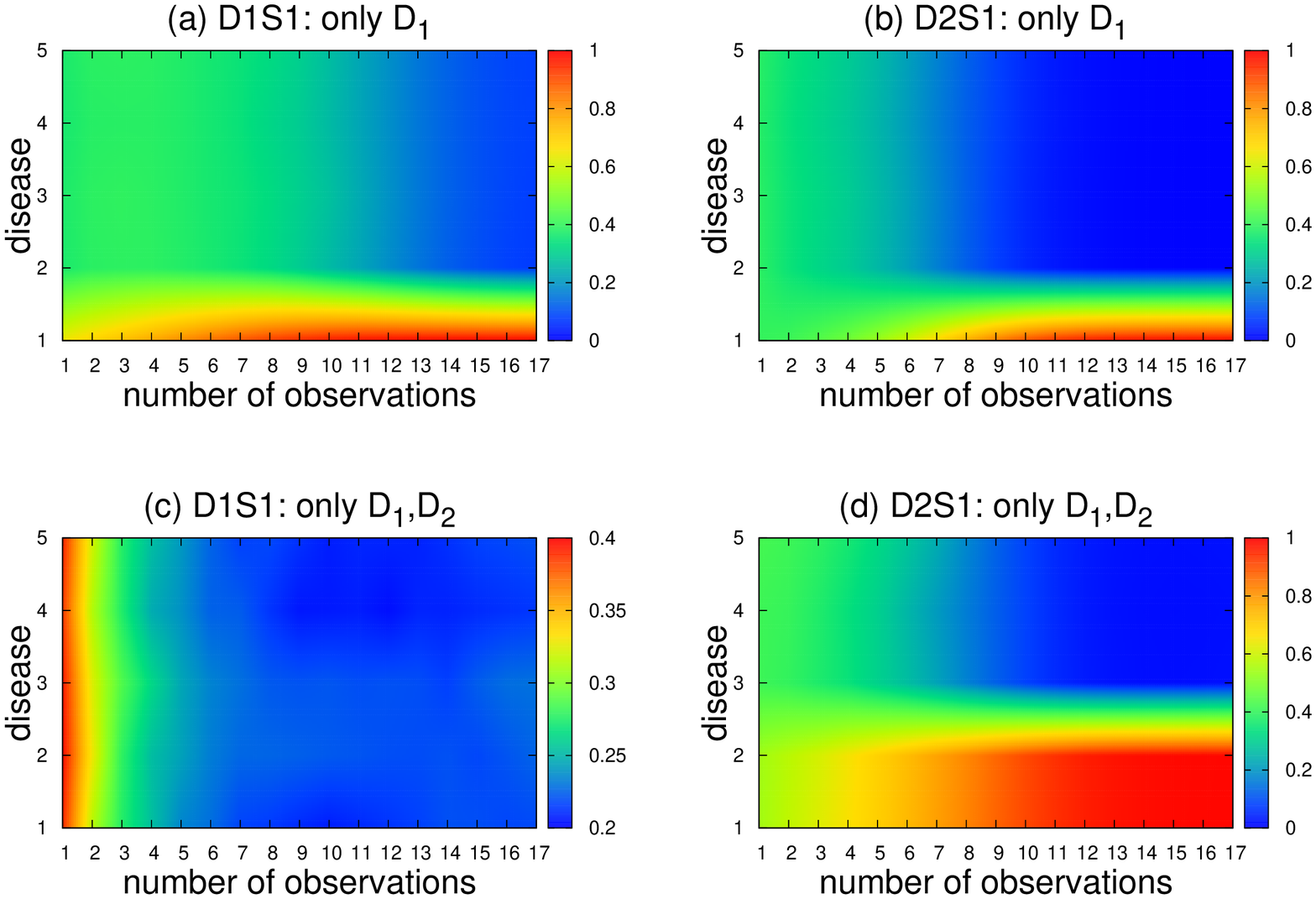} 
\caption{The inferred disease probabilities vs the number of observations for the exponential benchmark model. The data are for the cases in which only one (a,b) or two (c,d) diseases are present. The model parameters are obtained from the conditional marginals of the true model. There are $N_D=5$ diseases, $N_S=20$ signs, and the algorithm starts with $N_O=3$ observed signs for a randomly selected hypothesis $\mathbf{D}$. An unobserved sign is chosen for observation by the MP strategy using the inferred sign probabilities. The observed sign takes the true value given by $\mathbf{S}^*(\mathbf{D})$.
The data are results of averaging over $1000$ independent realizations of the true model and the observation process.}\label{Q2}
\end{figure}

\begin{figure}
\includegraphics[width=8cm]{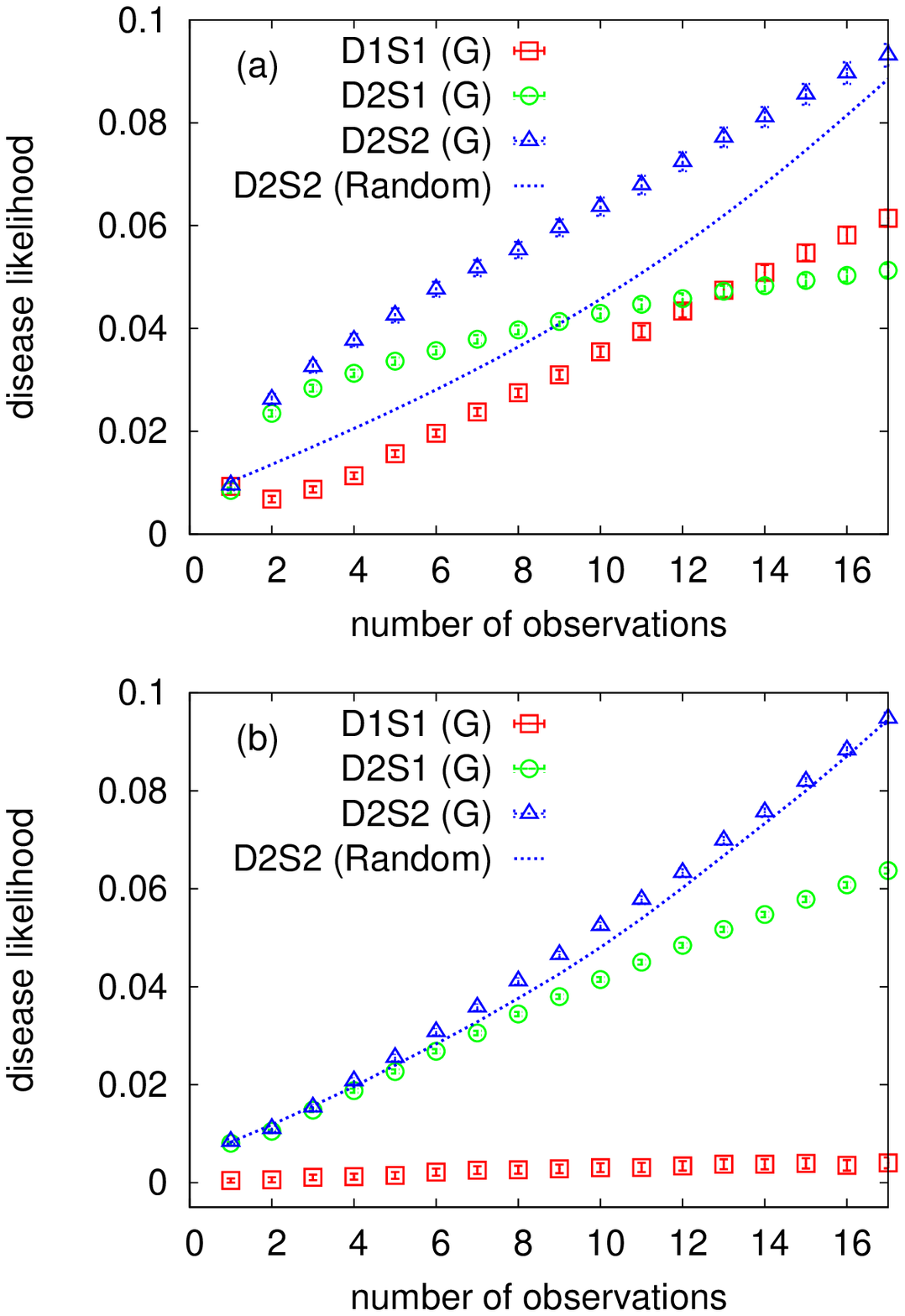} 
\caption{Overlap of the inferred disease marginals with the true hypothesis for the power-law benchmark models. The data are for the cases in which only one (a) or two (b) diseases are present. The model parameters are obtained from the conditional marginals of the true model. There are $N_D=5$ diseases, $N_S=20$ signs, and the algorithm starts with $N_O=3$ observed signs for a randomly selected hypothesis $\mathbf{D}$. An unobserved sign is chosen for observation by the greedy (G) strategy using the inferred sign probabilities. The observed sign takes the true value given by $\mathbf{S}^*(\mathbf{D})$. The data are results of averaging over $1000$ independent realizations of the true model and the observation process.}\label{Q3}
\end{figure}

Now, we consider a sequential process of $T$ decisions and real observations; at each step an unobserved sign is chosen for observation and the observed sign takes the true value, given by the symptoms $\mathbf{S}^*(\mathbf{D})$ of the true hypothesis $\mathbf{D}$. This hypothesis is selected randomly and uniformly at the beginning of the process. At each step, we compute the sign marginals $P(S_j)$ in the constructed models, and use the MP (or greedy) strategy to choose the next observation. Figure \ref{Q0} shows the RMS errors $\Delta_D$ and $\Delta_S$ in the marginal disease and sign probabilities, respectively, during such a process. Here, we are using the MP strategy and the errors are defined as follows,
\begin{align}
\Delta_D &\equiv \left(\frac{1}{N_D}\sum_{a}\left(P(D_a=1)-P_{true}(D_a=1)\right)^2 \right)^{1/2},\\
\Delta_S &\equiv \left(\frac{1}{N_U}\sum_{i\in U}\left(P(S_i=1)-P_{true}(S_i=1)\right)^2 \right)^{1/2},
\end{align}   
where $U$ is the subset of unobserved signs with $N_U=|U|$. Figure \ref{Q2} displays the inferred disease probabilities $P(D_a=1)$ in the same process for cases in which only one or two diseases are present. It is seen that both the $D1S1$ and $D2S1$ models end up with the right diagnosis when only one diseases is present ($D_1=1, D_2=D_3=\cdots=0$). But, the $D1S1$ model fails when there are two diseases in the hypothesis ($D_1=D_2=1, D_3=D_4=\cdots=0$), whereas the other model can still find the right diseases. However, we find that even the $D2S1$ model gets confused when there are more than two diseases in the hypothesis. Then, we would need to consider more sophisticated models with interactions involving more than two diseases.

To see the importance of sign-sign interactions, we consider another benchmark model given by the following conditional probability:
\begin{align}
P_{true}(\mathbf{S}|\mathbf{D})=\frac{1}{Z_{true}(\mathbf{D})}\left(\frac{1}{1+H(\mathbf{S},\mathbf{S}^*(\mathbf{D}))}\right).
\end{align} 
Given a disease hypothesis $\mathbf{D}$ and the associated symptoms $\mathbf{S}^*(\mathbf{D})$, we start from a few number of randomly chosen observed sings from the set of symptoms $\mathbf{S}^*(\mathbf{D})$. At each step, we compute the inferred sign probabilities $P(S_i)$ and use the greedy algorithm to choose an unobserved sign for observation. The observed sign at each time step takes the true value given by $\mathbf{S}^*(\mathbf{D})$. Figure \ref{Q3} shows the overlap of the inferred probabilities $P(D_a=1)$ with the true hypothesis for cases in which only one or two diseases are present. For comparison, we also show the results obtained by the random strategy. The figure shows that we can obtain higher overlaps between the inferred disease marginals and the true hypothesis by considering the sign-sign interactions. Nevertheless, we find that even the $D2S2$ model fails in cases with more than two diseases.

\subsection{Case II}\label{app-32}
Here, the model parameters are chosen randomly and uniformly in symmetric intervals of real numbers; e.g., $K_i^{a,ab} \in (-1,+1), K_{ij}^{a,ab} \in (-1,+1)/\sqrt{N_S}$. The leaky couplings are set to $K_i^0=-1$, corresponding to small sign probabilities $P(S_i=1|\mathrm{nodisease}) \simeq 0.05$. We assume that all the possible one-disease and two-disease interaction factors are present in the models. Moreover, inside each factor we have all the possible two-sign interactions in addition to the local sign fields; that is why we scale the two-sign couplings $K_{ij}^{a,ab}$  with the inverse square root of the number of sign variables. The prior disease probabilities are uniform $P_{0}(D_a=1)=0.5$.

\begin{figure}
\includegraphics[width=14cm]{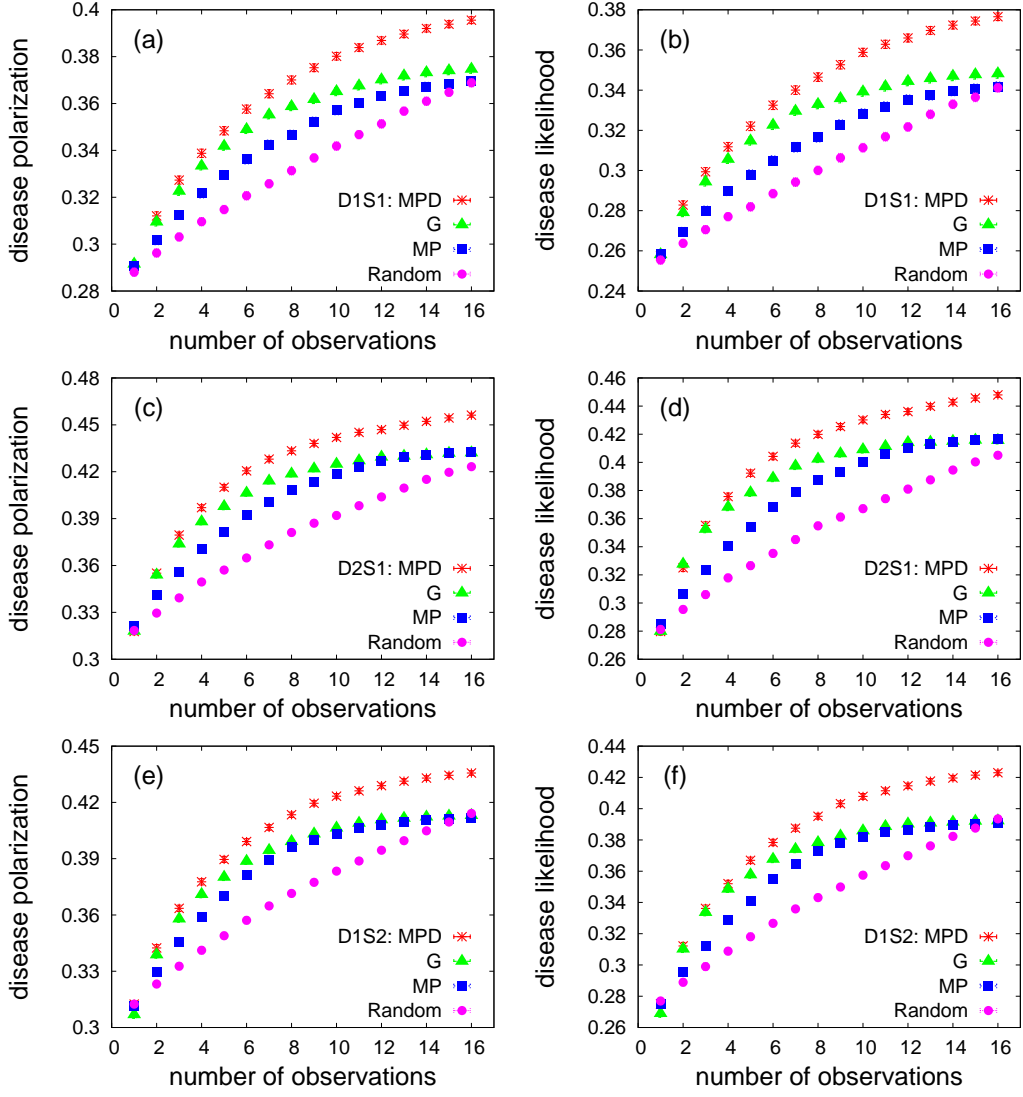} 
\caption{Diagnostic performance of the models vs the number of observations for different strategies. The exact disease-polarization (DP) and  disease-likelihood (DL) obtained by the greedy, MP, MPD, and random strategies in the one-disease-one-sign ($D1S1$, (a)-(b)), two-disease-one-sign ($D2S1$, (c)-(d)), and one-disease-two-sign ($D1S2$, (e)-(f)) models. There are  $N_D=5$ diseases, $N_S=20$ signs, and the algorithm starts with $N_O=4$ observed signs with positive values. The coupling constants are iid random numbers distributed uniformly in the specified intervals: $K_i^0=-1, K_i^{a,ab} \in (-1,+1), K_{ij}^{a,ab} \in (-1,+1)/\sqrt{N_S}$. The data are results of averaging over $1000$ independent realizations of the model parameters.}\label{Dtt}
\end{figure}

\begin{figure}
\includegraphics[width=14cm]{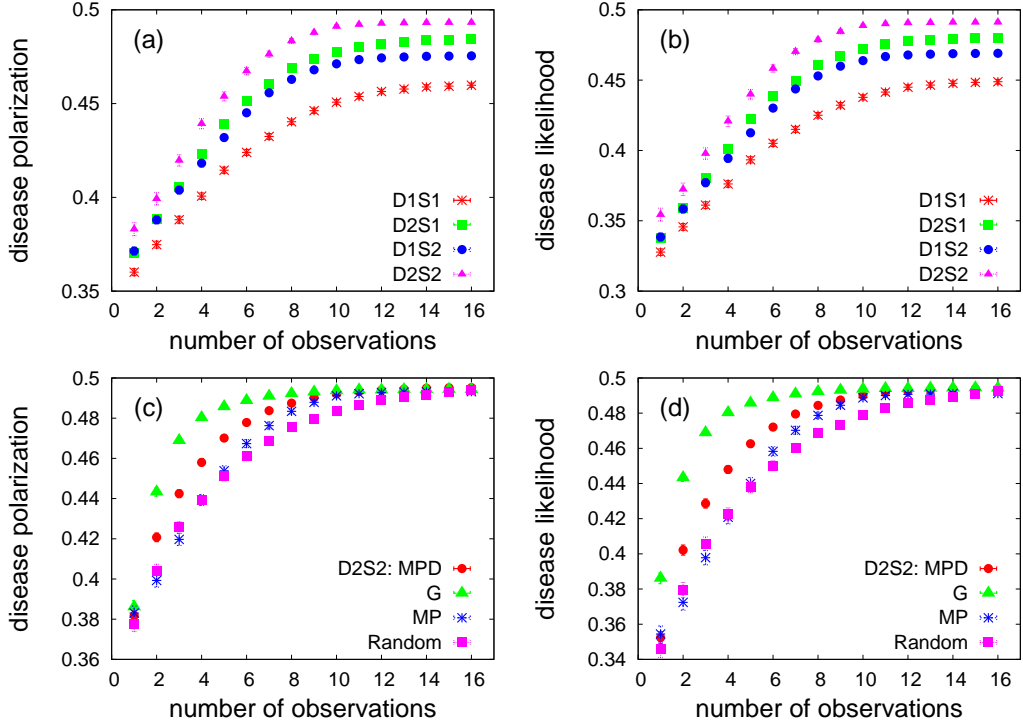} 
\caption{Diagnostic performance of the models vs the number of observations for larger couplings. (top) The exact disease polarization (a) and disease likelihood (b) obtained by the MP strategy in the one-disease-one-sign ($D1S1$), two-disease-one-sign ($D2S1$), one-disease-two-sign ($D1S2$), and two-disease-two-sign ($D2S2$) models. There are  $N_D=5$ diseases, $N_S=20$ signs, and the algorithm starts with $N_O=4$ observed signs with positive values. 
The couplings in the interaction factors are iid random numbers distributed uniformly in the specified intervals: $K_i^0=-1, K_i^{a,ab} \in (-2,+2), K_{ij}^{a,ab} \in (-2,+2)/\sqrt{N_S}$. (bottom) Comparing the exact polarization (c) and likelihood (d) of the diseases obtained by the greedy, MP, MPD, and random strategies, for the $D2S2$ model. The data are results of averaging over $1000$ independent realizations of the model parameters.}\label{Dt-K12}
\end{figure}

\begin{figure}
\includegraphics[width=8cm]{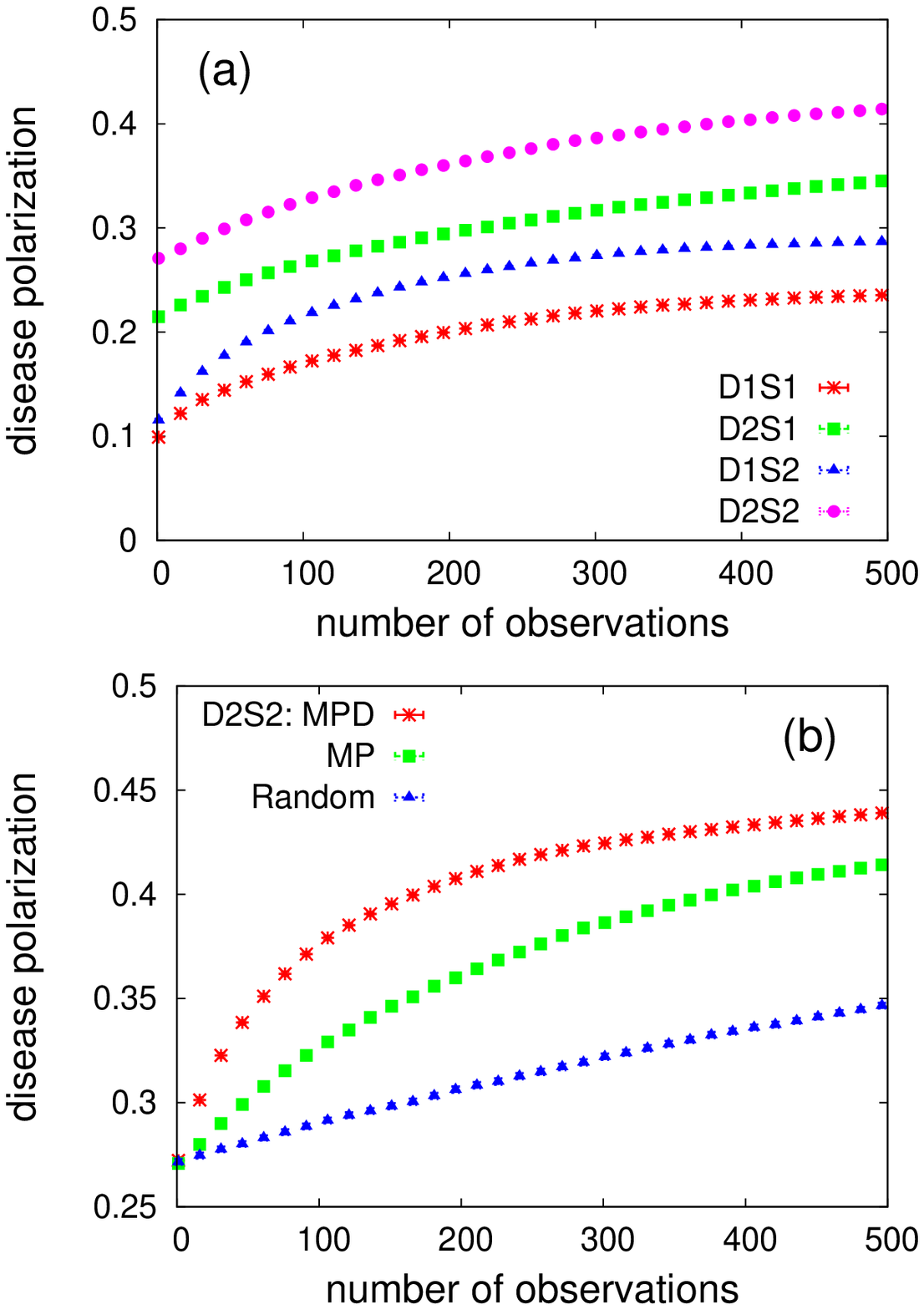} 
\caption{Diagnostic performance of the models vs the number of observations for a larger number of sign/disease variables. (a) The approximate disease-polarization (DP) obtained by the MP strategy in the one-disease-one-sign ($D1S1$), two-disease-one-sign ($D2S1$), one-disease-two-sign ($D1S2$), and two-disease-two-sign ($D2S2$) models. Here we have  $N_D=200$ diseases, $N_S=2000$ signs, and the algorithm starts with $N_O=50$ observed signs with positive values. 
The graph parameters are: $M_a=200, M_{ab}=0, k_a=8$ in the $D1S1$ and $D1S2$ models, and $M_a=200, M_{ab}=1000, k_a=8, k_{ab}=4$ in the $D2S1$ and $D2S2$ models. The interaction factors in the $D1S2$ and $D2S2$ models include all the possible two-sign interactions in addition to the local sign fields. The couplings in the interaction factors are iid random numbers distributed uniformly in the specified intervals: $K_i^0=-1, K_i^{a,ab} \in (-1,+1), K_{ij}^{a,ab} \in (-1,+1)/\sqrt{k_{a,ab}}$. (b) Comparing the approximate polarization of the diseases obtained by the MP, MPD, and random strategies, for the $D2S2$ model. The data are results of averaging over $100$ independent realizations of the model parameters.}\label{Dt-BP}
\end{figure}

We start from a small number of randomly selected observed signs with random values. Then, at each step, we choose an unobserved sign for observation with stochastic findings sampled from the associated probability distributions. The process is repeated $1000$ times for different realizations of the couplings and findings of the observations to obtain the average polarization and likelihood of diseases. For reference, we also show the results that were obtained by a random strategy, where an unobserved sign is chosen randomly and uniformly for observation. In Fig. \ref{Dtt} we see that the average performance of the greedy strategy changes from that of the MPD strategy for small number of observations to that of the MP strategy for larger number of observations. 

However, as Fig. \ref{Dt-K12} shows, for strongly correlated models with much larger higher-order interactions, e.g., $K_i^0=-1$ and $K_i^{a,ab} \in (-2,+2), K_{ij}^{a,ab} \in (-2,+2)/\sqrt{N_S}$, the difference between the performances of the most positive (MP) strategy and the random strategy diminishes. Furthermore, here the greedy strategy performs better than the MP and MPD strategies. The fact that such a heuristic algorithm does not work well in this regime was indeed expected; it is this type of finding that makes strongly interacting systems interesting, and in need for more elaborate methods and algorithms.     

Finally, Fig. \ref{Dt-BP} displays the average polarization of disease probabilities obtained by the approximate inference algorithm (BP) for much larger number of sign/disease variables. Here, the maximum likelihood hypothesis ($\mathbf{D}_{ML}$ in the MPD strategy) is simply approximated by the most probable diseases suggested by the marginal disease probabilities. That is, we take $D_a=0$ if $P(D_a=0) > 1/2$, otherwise $D_a=1$.

\end{document}